\documentclass[twocolumn,aps,showpacs]{revtex4}
\def  \be   {\begin{equation}}
\def  \ee   {\end{equation}}
\def  \beq  {\begin{eqnarray}}
\def  \eeq  {\end{eqnarray}}
\def  \etal   {{\em et~al.~}}

\begin{document}

\title{Magnetic reconnection with anomalous resistivity in
two-and-a-half dimensions~I: Quasi-stationary case}

\author{Leonid M.~Malyshkin}
\email{leonmal@flash.uchicago.edu}
\author{Timur Linde}
\email{linde@flash.uchicago.edu}
\affiliation{Department of Astronomy, The University of Chicago --
The Center for Magnetic Self-Organization (CMSO), Chicago, IL 60637.
}

\author{Russell M.~Kulsrud}
\email{rmk@pppl.gov}
\affiliation{
Princeton Plasma Physics Laboratory -- The Center for 
Magnetic Self-Organization (CMSO), Princeton, NJ 08543. 
}

\date{\today}

\begin{abstract}
In this paper quasi-stationary, two-and-a-half-dimensional magnetic
reconnection is studied in the framework of incompressible resistive
magnetohydrodynamics (MHD). A new theoretical approach for calculation
of the reconnection rate is presented. This approach is based on local
analytical derivations in a thin reconnection layer, and it is
applicable to the case when resistivity is anomalous and is an
arbitrary function of the electric current and the spatial
coordinates. It is found that a quasi-stationary reconnection rate is
fully determined by a particular functional form of the anomalous
resistivity and by the local configuration of the magnetic field just
outside the reconnection layer. It is also found that in the special
case of constant resistivity reconnection is Sweet-Parker and not
Petschek. 
\end{abstract}

\pacs{52.30.Cv, 52.35.Vd, 52.65.Kj}

\maketitle


\section{Introduction}
\label{INTRODUCTION}

Magnetic reconnection is the physical process of breaking and
rearrangement of magnetic field lines, which changes the topology of
the field. It is one of the most fundamental processes of plasma
physics and is believed to be at the core of many dynamic phenomena in
laboratory experiments and in cosmic space. Unfortunately, in spite of
being so important, magnetic reconnection is still relatively
poorly understood from the theoretical point of view. The reason is
that plasmas usually have very high temperatures and low densities. In
such plasmas, the Spitzer resistivity is extremely small
and magnetic fields are almost perfectly frozen into cosmic
plasmas. As a result, simple theoretical models, such as the
Sweet-Parker reconnection model~\cite{sweet_1958,parker_1963} predict
that the magnetic reconnection processes should be extremely slow and
insignificant throughout the universe. On the other hand,
astrophysical observations show magnetic reconnection tends to be fast
and is likely to be the primary driver of many highly energetic cosmic
processes, such as solar flares and geomagnetic storms. This
contradiction between theoretical estimates and astrophysical
observations triggered multiple attempts to build a theoretical model
of fast magnetic reconnection.

First, in 1964 Petschek proposed a fast reconnection
model~\cite{petschek_1964}, in which fast reconnection is achieved by
introducing switch-off magnetohydrodynamic (MHD) shocks attached to
the ends of the reconnection layer in the downstream regions and by
choosing the reconnection layer length to be equal to its minimal
possible value under the condition of no significant disruption to the
plasma flow. However, later numerical simulations and theoretical
derivations did not confirm the Petschek theoretical picture for the
geometry of a reconnection
layer~\cite{biskamp_1986,kulsrud_2001,uzdensky_2000}. 
Second, numerical simulations studies of anomalous magnetic
reconnection, for which resistivity is enhanced locally in
the reconnection layer, were pioneered by Ugai and
Tsuda~\cite{ugai_1977,tsuda_1977}, by Hayashi and
Sato~\cite{hayashi_1978,sato_1979}, and by
Scholer~\cite{scholer_1989}. Third, Lazarian and
Vishniac proposed that fast reconnection can
occur in turbulent plasmas~\cite{lazarian_1999}, although the
back-reaction of magnetic fields can slow down reconnection in
this case~\cite{kim_2001}.  
Finally, recently there have been number of attempts to explain the
fast magnetic reconnection by considering non-MHD 
effects~\cite{biskamp_1995,bhattacharjee_2003,craig_2005,drake_2003,
hanasz_2003,heitsch_2003,rogers_2003,shay_2001}.

Most previous theoretical and numerical studies concentrated on
reconnection processes in two-dimensions or in
``two-and-a-half-dimensions''. The later is the term used for a
problem in which physical scalars and all three components of physical
vectors depend only on two spatial coordinates (e.g.~$x$ and $y$) and
are independent of the third coordinate ($z$).

\begin{figure}[t]
\vspace{7.4truecm}
\includegraphics{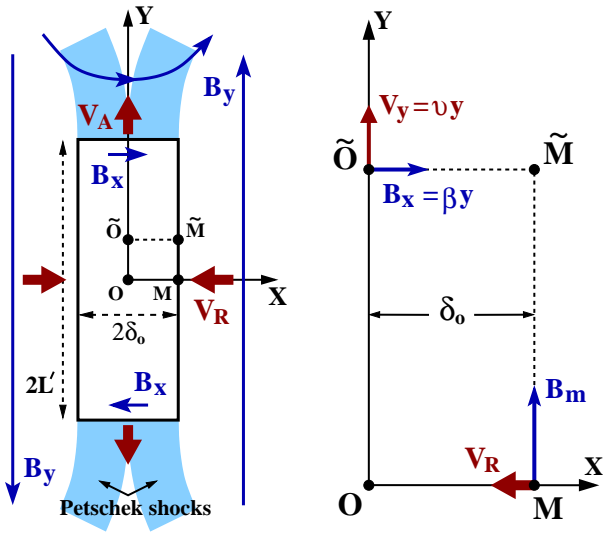}
\caption{(Color online) The geometrical configuration of the classical
Sweet-Parker-Petschek reconnection layer is shown in the left
plot. Petschek shocks exist only if the reconnection is considerably
faster than the Sweet-Parker reconnection rate. The right 
plot shows an enlarged picture of the central layer region.
} 
\label{FIG_RECONNECTION_LAYER}
\end{figure}

In this paper we consider two-and-a-half-dimensional magnetic
reconnection with anomalous resistivity in the classical
Sweet-Parker-Petschek reconnection layer, which is shown in the left
plot in Fig.~\ref{FIG_RECONNECTION_LAYER}. The reconnection layer is
in the x-y plane with the y-axis being along the layer and the x-axis
being perpendicular to the layer. The length of the layer is equal to
$2L'$. Note that $L'$ is approximately equal to or smaller than the
global magnetic field scale, which we denote as $L$. The thickness of
the classical reconnection layer, $2\delta_o$, is much smaller than
its length, i.e.~$2\delta_o\ll2L'$. The classical Sweet-Parker-Petschek
reconnection layer is assumed to possess a point symmetry with respect
to its geometric center point O and reflection symmetries with respect
to the axes $x$ and $y$ (refer to Fig.~\ref{FIG_RECONNECTION_LAYER}).
Thus, for example, the x- and y-components of the plasma velocities 
${\bf V}$ and of the magnetic field ${\bf B}$ have the following
simple symmetries:  
$V_x(\pm x,\mp y)=\pm V_x(x,y)$, $V_y(\pm x,\mp y)=\mp V_y(x,y)$, 
$B_x(\pm x,\mp y)=\mp B_x(x,y)$ and $B_y(\pm x,\mp y)=\pm B_y(x,y)$. 
There might be a pair of Petschek shocks attached to each of the 
two reconnection layer ends in the downstream regions (see
Fig.~\ref{FIG_RECONNECTION_LAYER}). Because of the MHD jump 
conditions on the Petschek shocks, the presence of these shocks
requires the presence of a significant perpendicular magnetic field
$B_x$ at the reconnection layer
ends~\cite{kulsrud_2001,kulsrud_2005}. The plasma outflow velocity
from the reconnection layer is approximately equal to the Alfven
velocity $V_A$ (if the plasma viscosity is not very large). The plasma
inflow velocity $V_R$ outside of the reconnection layer, at point M in
Fig.~\ref{FIG_RECONNECTION_LAYER}, is much smaller than the outflow
velocity, $V_R\ll V_A$. Finally, the magnetic field outside the
reconnection layer is mostly in the direction of the layer (i.e.~in
the y-axis direction).

The problem of quasi-stationary anomalous magnetic reconnection in the
classical Sweet-Parker-Petschek reconnection layer was recently
theoretically addressed by Kulsrud~\cite{kulsrud_2001} for the special
case of zero guide field ($B_z=0$), zero plasma viscosity and
anomalous resistivity that is a piecewise linear function of the
electric current [see Eq.~(\ref{LINEAR_RESISTIVITY})]. There are
two major results of the Kulsrud anomalous reconnection model. 
First, in the case of the Petschek geometry of the reconnection layer
and a constant resistivity, one has to calculate the half-length of
the reconnection layer $L'$ from the MHD equations and the jump
conditions on the Petschek shocks, instead of treating $L'$ as a free
parameter (as Petschek did). When the layer half-length $L'$ is
calculated correctly, it turns out to be approximately equal to the
global magnetic field scale, $L'\approx L$. In this case the Petschek
reconnection reduces to the slow Sweet-Parker
reconnection~\cite{kulsrud_2001,kulsrud_2005}. This theoretical result
is in agreement with the results of numerical simulations of
two-dimensional reconnection with constant
resistivity~\cite{biskamp_1986,uzdensky_2000}. The second major result
of the Kulsrud reconnection model is that in the case when resistivity
is anomalous and enhanced (e.g. by plasma instabilities), the
reconnection rate becomes considerably faster than the Sweet-Parker
rate. 

In this paper we develop and use a new theoretical approach for 
calculation of the reconnection rate in the case of anomalous
resistivity. This approach is based on application of the MHD
equations in a small region, which is localized at the geometric
center of a thin reconnection layer (point O in
Fig.~\ref{FIG_RECONNECTION_LAYER}) and has a size of order of the
layer half-thickness $\delta_o$. It turns out that by using local
analytical calculations in a thin reconnection layer, we can derive an
accurate and rather precise estimate for the reconnection rate. In
particular, we find the interesting and important result that a
quasi-stationary reconnection rate is fully determined by the
anomalous resistivity function and by the magnetic field configuration
just outside the reconnection layer (at point M in
Fig.~\ref{FIG_RECONNECTION_LAYER}). The underlying physical
foundations of our analytical derivations are, of course, the same as
those previously used by
others~\cite{sweet_1958,parker_1963,kulsrud_2001}. However, our
theoretical and computational approach is somewhat different from the
conventional approach to the reconnection problem, and we explain the
difference in the next section. There are three main benefits of our
novel theoretical approach to magnetic reconnection. First, our
approach allows us to extend the results for anomalous reconnection
obtained by Kulsrud in 2001~\cite{kulsrud_2001} to a general anomalous
reconnection case, when the guide field and plasma viscosity are
arbitrary and anomalous resistivity $\eta$ is an arbitrary function of
the electric current and the two spatial coordinates. Second, it gives
a new important insight into the reconnection problem, such as
dependence of the reconnection rate on magnetic field configuration
just outside the reconnection layer. Third, our approach, based on
local calculations, is applicable to cases when there is no
well-defined global magnetic field structure, such as the
case of multiple current sheets in a turbulent plasma. In our
calculations we make only a few assumptions, which are described in
detail in the beginning of Sec.~\ref{NEW_MODEL}. 

This paper is organized as follows. Because our derivations
are rather complicated in the general case of reconnection with
anomalous resistivity, we find it useful and instructive to consider
the Sweet-Parker reconnection first and to compare our theoretical
approach to the classical Sweet-Parker calculations in the next
section. In Sec.~\ref{NEW_MODEL}, we derive our general equations for
magnetic reconnection with anomalous resistivity, including the
equation for the reconnection rate. In Sec.~\ref{RECONNECTION_LIMITS}
we consider and analyze different special cases of magnetic
reconnection, in which our equations simplify and become
easier for analysis and comparison to the previous theoretical and
simulation results. In Sec.~\ref{SIMULATIONS} we present the results
of our numerical simulations of unforced anomalous magnetic
reconnection. These simulations are intended for a demonstration of
the predictions of our theoretical reconnection model for the special
case of the piecewise linear resistivity function that was considered
by Kulsrud~\cite{kulsrud_2001}. Finally, in Sec.~\ref{CONCLUSIONS} we
give our conclusions and discuss our results. Some derivations are
given in the appendices of the paper.


\section{The Sweet-Parker model of magnetic reconnection}
\label{SWEET_PARKER_MODEL}

For simplicity and brevity, hereafter in the paper
for all electromagnetic variables we use physical units in which
the speed of light and four times $\pi$ are replaced unity,
$c=1$ and $4\pi=1$. To rewrite our equations in the standard CGS
units, one needs to make the following substitutions for
electromagnetic variables in the equations: magnetic field 
${\bf B}\rightarrow {\bf B}/\sqrt{4\pi}$, electric field 
${\bf E}\rightarrow c{\bf E}/\sqrt{4\pi}$, electric current 
${\bf j}\rightarrow (\sqrt{4\pi}/c){\bf j}$, resistivity 
$\eta\rightarrow\eta$ (does not change), magnetic field vector 
potential ${\bf A}\rightarrow {\bf A}/\sqrt{4\pi}$.

In this section we consider Sweet-Parker
reconnection~\cite{sweet_1958,parker_1963}. We assume that
resistivity is constant, $\eta\equiv{\rm const}=\eta_o$, the plasma
viscosity is zero, the guide field is zero ($B_z=0$), and the geometry
of the reconnection layer is the classical Sweet-Parker geometry with
the layer half-thickness $\delta_o\ll L$ and the layer half-length
$L'=L$, as shown in Fig.~\ref{FIG_RECONNECTION_LAYER}. The purpose
of this section is to introduce and explain our new theoretical
approach to calculation of the reconnection rate and to compare it
with the classical Sweet-Parker calculations. For this purpose, we
first present the classical, conventional derivation of the
Sweet-Parker formula for the reconnection rate, and afterward we
present our new derivation of the formula and explain the difference
between the two derivations.

In the case of a quasi-stationary reconnection the
magnetic field in the reconnection region changes slowly in time, 
$\partial{\bf B}/\partial t\approx0$. Therefore, in the
two-and-a-half-dimensional geometry ($\partial/\partial z\equiv 0$)
the Faraday's law equation 
$-{\bf\nabla}\times{\bf E}=\partial{\bf B}/\partial t\approx0$ results
in the z-component of the electric field being constant in the
reconnection region, ${\bf\nabla}E_z=0$ and $E_z=E_z(t)$ is a function
of time only. On the other hand, in two-and-a-half dimensions
the $x$- and $y$-components of Faraday's law equation 
$\partial{\bf B}/\partial t=-{\bf\nabla}\times{\bf E}=
-{\bf\nabla}\times(\eta{\bf j}-[{\bf V}\times{\bf B}])$ reduce to 
the following equation for the z-component of the magnetic vector
potential ${\bf A}$: 
\beq
-E_z(t)&=&\partial A_z/\partial t=
-({\bf V}\cdot{\bf\nabla})A_z+\eta{\bf\nabla}^2 A_z
\nonumber
\\
{}&=&V_xB_y-V_yB_x-\eta j_z.
\label{MHD_EQUATION}
\eeq
(The resistivity $\eta\equiv{\rm const}=\eta_o$ is constant in the
Sweet-Parker reconnection case, but equation~(\ref{MHD_EQUATION}) is
general and is valid even if resistivity $\eta$ is anomalous and
non-constant.)

Now, the left- and right-hand-sides of equation~(\ref{MHD_EQUATION})
are constant in space. Therefore, the right-hand-side of
equation~(\ref{MHD_EQUATION}) is constant across the reconnection
layer (i.e.~along the x-axis). We equate its values at points O
and M, which are on the x-axis and are shown in
Fig.~\ref{FIG_RECONNECTION_LAYER}. As a result, we immediately 
obtain 
\beq
\eta_oj_o=V_RB_m,
\label{FIRST_MHD}
\eeq
where we use the following notations: Point O is the geometric
center of the reconnection layer, where the z-component of the current
is $j_o=j_z(x=0,y=0)$ and the plasma velocity is zero. Point M is a
point on the x-axis just outside the reconnection layer, where the
resistivity term can be neglected in equation~(\ref{MHD_EQUATION}),
see Fig.~\ref{FIG_RECONNECTION_LAYER}. We also use the notations
$B_y=B_m$ and $V_x=-V_R$ for the y-component of the field and
x-component of the plasma velocity at point M, and take the
reconnection velocity $V_R$ positive. Note that $V_y=B_x=0$ at point M
because of the symmetry of the problem with respect to the
x-axis. Next, we estimate current $j_o$ at the central point O of the
reconnection layer as 
\beq
j_o\approx B_m/\delta_o,
\label{J_O}
\eeq
where $\delta_o$ is the reconnection layer half-thickness, 
and we use Ampere's law,
$j_o=(\partial B_y/\partial x)_o-(\partial B_x/\partial y)_o\approx
(\partial B_y/\partial x)_o\approx B_m/\delta_o$, in which we drop
the $(\partial B_x/\partial y)_o$ term because the reconnection layer
is 
thin~\footnote{ 
Otherwise, if 
$|\partial B_x/\partial y|_o\approx|\partial B_y/\partial x|_o$, 
then obviously the length of the layer would be of the order of its
thickness.
}.

Next, consider the x- and y-components of the equation of plasma
motion. We will see below that the Sweet-Parker reconnection is
slow, $V_R\ll V_A\equiv B_m/\sqrt{\rho}$. Therefore, in the equation
of plasma motion along the x-axis (i.e.~across the reconnection layer)
the inertial term can be neglected, and this equation becomes the
force balance equation, $(\partial/\partial x)(P+B^2/2)=0$, resulting
in $P_o=P_m+B_m^2/2$, where $P_o$ and $P_m$ are the values of the
plasma pressure at points O and M. Here we use the fact that the
magnetic field is zero at the central point O because of the symmetry
of the problem. As far as the equation of plasma motion along the
y-axis (i.e~along the layer) is concerned, the y-components of the
magnetic tension force and the pressure gradient force are
approximately equal in the Sweet-Parker reconnection case. Indeed, on
the y-axis the tension force can be estimated as 
$({\bf B}\cdot{\bf\nabla})B_y=B_x\times(\partial B_y/\partial x)
\approx(\delta_o/L)B_m\times(B_m/\delta_o)=B_m^2/L\approx(P_o-P_m)/L
\approx-\partial P/\partial y$. Here we estimate the x-component of
the field as $B_x\approx(\delta_o/L)B_m$ because $B_x$ is produced by
the rotation of the $B_y$ field component in the reconnection
layer~\cite{kulsrud_2001}, and in the Sweet-Parker model it is assumed
that the downstream pressure is equal to the upstream pressure
$P_m$ (see~\cite{sweet_1958,parker_1963}). The pressure and magnetic
tension forces accelerate plasma along the reconnection layer
(i.e.~along the y-axis) up to the downstream velocity $V_{out}$, which
can be estimated from the energy conservation equation  
\beq
(1/2)\rho V_{out}^2\approx B_m^2/2
\;\;\Rightarrow\;\; V_{out}\approx V_A\equiv B_m/\sqrt{\rho}.
\label{SP_ENERGY}
\eeq
This equation means that the work done by the pressure and
magnetic tension forces along the entire reconnection layer is equal
to the kinetic energy of the plasma in the downstream regions. 

Finally, in the Sweet-Parker reconnection model the plasma is assumed
to be incompressible. Therefore, the mass conservation condition for
the entire reconnection layer results in 
\beq
LV_R\approx\delta_oV_{out},
\label{SP_MASS}
\eeq
where $V_{out}$, the velocity of plasma outflow in the downstream
regions, is given by equation~(\ref{SP_ENERGY}). Using
equations~(\ref{FIRST_MHD})-(\ref{SP_MASS}), we obtain the 
formula for the Sweet-Parker reconnection
velocity~\cite{sweet_1958,parker_1963},
\beq
V_R\approx V_A(\eta_o/V_AL)^{1/2}.
\label{SP_SPEED}
\eeq

Equations~(\ref{FIRST_MHD})-(\ref{SP_SPEED}) are the classical
Sweet-Parker equations. Note that
equations~(\ref{FIRST_MHD}),~(\ref{J_O}) and~(\ref{SP_SPEED}) are
local, in the sense that they are written for a small region of space,
which is localized at the geometric center of a thin reconnection
layer (point O in Fig.~\ref{FIG_RECONNECTION_LAYER}) and has a size
of order of the layer half-thickness $\delta_o$. All physical
quantities that enter these three equations are defined in this small
region of space. At the same time equations~(\ref{SP_ENERGY})
and~(\ref{SP_MASS}) are global, in the sense that they result from
consideration of the entire reconnection layer and they include the
plasma outflow velocity $V_{\rm out}$, which is a physical quantity in
the downstream regions at the ends of the reconnection layer. In our
new theoretical approach to the calculation of the reconnection rate
we intend to use only local equations. Our intent and our derivations
will be justified by the new and important results that we obtain in
the next section of this paper and discuss in
Sec.~\ref{CONCLUSIONS}. At present, let us explain our theoretical
approach for the simple case of Sweet-Parker reconnection that we
consider in this section. 

In the derivation of our theoretical model for magnetic reconnection
we keep equations~(\ref{FIRST_MHD}) and~(\ref{J_O}) 
unchanged because these equations are local. However, we rewrite
equations~(\ref{SP_ENERGY}) and~(\ref{SP_MASS}) because they are
global. To rewrite these two global equations in a local form, we
consider a point ${\tilde{\rm O}}=(x=0,y={\tilde y})$ that is located
on the y-axis in an infinitesimal vicinity of the reconnection layer
central point O (see Fig.~\ref{FIG_RECONNECTION_LAYER}) and has an
infinitesimally small value of its y-coordinate ${\tilde y}$. Since
${\tilde y}\to+0$, along the O${\tilde{\rm O}}$ interval we can use
the up-to-the-first-order Taylor expansions  $B_x(0,y)=\beta y$,
$V_y(0,y)=\upsilon y$ and $j_z(0,y)=j_o$ for the values of the
perpendicular magnetic field $B_x$, plasma velocity $V_y$ and
z-component of the current $j_z$. These expansions are along the
y-axis, and, of course, $\beta\equiv(\partial B_x/\partial y)_o$ and 
$\upsilon\equiv(\partial V_y/\partial y)_o$ are the first-order
partial derivatives of $B_x$ and $V_y$ at point O (note that
$(\partial J_z/\partial y)_o=0$). Now equation~(\ref{SP_ENERGY}) for
the plasma acceleration along the y-axis can easily be rewritten in a
local form at point ${\tilde{\rm O}}$ as
\beq
\frac{1}{2}\rho(\upsilon{\tilde y})^2 &\approx&
\int_0^{\tilde y}B_x(0,y)j_z(0,y)dy=
\int_0^{\tilde y}\beta y j_o dy
\nonumber\\
{}&=&\beta j_o{\tilde y}^2/2
\quad\Rightarrow\quad
\rho\upsilon^2\approx\beta j_o,
\label{SP_LOCAL_ENERGY}
\eeq
where the left-hand-side of this equation is the plasma kinetic energy
at point ${\tilde{\rm O}}$. On the right-hand-side of
equation~(\ref{SP_LOCAL_ENERGY}) we keep only the magnetic tension
force for plasma acceleration because, as we found above, in the
Sweet-Parker reconnection case the y-components of the magnetic
tension and pressure gradient forces are approximately equal to each
other (note that in our general calculations in the next section
we will take all forces into account). Next, equation~(\ref{SP_MASS})
for the mass conservation of an incompressible plasma can easily be
rewritten in the local form inside the area 
OM${\tilde{\rm M}}$${\tilde{\rm O}}$ shown in
Fig.~\ref{FIG_RECONNECTION_LAYER} as 
${\tilde y}V_R\approx\delta_oV_y(0,{\tilde y})=
\delta_o\upsilon{\tilde y}$. Thus, we have
\beq
\upsilon\approx V_R/\delta_o, \quad 
\mbox{where~}\upsilon\equiv(\partial V_y/\partial y)_o=
-(\partial V_x/\partial x)_o.
\label{UPSILON_ESTIMATE}
\eeq
This equation can also be viewed as the first order Taylor expansion
of $V_x$ along the x-axis (i.e.~across the reconnection layer), 
$V_R\equiv-V_x(\delta_o,0)\approx-(\partial V_x/\partial x)_o\delta_o
=(\partial V_y/\partial y)_o\delta_o=\upsilon\delta_o$, where we use
the plasma incompressibility condition 
$\partial V_x/\partial x+\partial V_y/\partial y=0$.

Note that our first and second equations are local and are exactly the
same as equations~(\ref{FIRST_MHD}) and~(\ref{J_O}) in the
Sweet-Parker reconnection model. Our third and fourth
equations~(\ref{SP_LOCAL_ENERGY}) and~(\ref{UPSILON_ESTIMATE}) are
also local and are the analogues of the two global Sweet-Parker
equations~(\ref{SP_ENERGY}) and~(\ref{SP_MASS}). Using local
equations instead of global ones is the first major difference
between the Sweet-Parker and our theoretical models.
However, in our theoretical model we have an additional unknown
parameter $\beta\equiv(\partial B_x/\partial y)_o$, which does not
directly enter the classical Sweet-Parker reconnection model. As a
result, we need one additional equation to be able to calculate the
reconnection rate under the framework of our local model. This 
additional equation comes from the condition that the
right-hand-side of equation~(\ref{MHD_EQUATION}) is constant not only
across the reconnection layer, but also along the layer that is
along the y-axis. This condition is not explicitly used in the
Sweet-Parker model, but it is used in our model, and this is the
second major difference between the two models. To derive the
additional equation, used in our reconnection model, we differentiate
the left- and right-hand-sides of equation~(\ref{MHD_EQUATION}) along
the y-axis (i.e.~along the reconnection layer). The first order
partial derivatives $\partial/\partial y$ are identically zero because
of the symmetry of the problem with respect to the x-axis. The
second order partial derivatives $\partial^2/\partial y^2$ of the
left- and right-hand-sides of equation~(\ref{MHD_EQUATION}) result in
\beq
0&=&-2(\partial V_y/\partial y)_o(\partial B_x/\partial y)_o
-\eta_o(\partial^2 j_z/\partial y^2)_o
\nonumber\\
{}&\approx&-2\upsilon\beta+2\eta_oj_o/L^2,
\label{SP_THIRD_MHD}
\eeq
where we use the fact that $V_x=B_y=0$ on the y-axis, we use our
definitions $\beta\equiv(\partial B_x/\partial y)_o$ and 
$\upsilon\equiv(\partial V_y/\partial y)_o$, and we estimate the
second derivative of the current $j_z$  as
$(\partial^2j_z/\partial y^2)_o\approx-2j_o/L^2$. Now all five 
equations~(\ref{FIRST_MHD}),~(\ref{J_O}),~(\ref{SP_LOCAL_ENERGY})-(\ref{SP_THIRD_MHD})
are local. Combining them together, we easily obtain the Sweet-Parker
formula for the reconnection velocity, given by
equation~(\ref{SP_SPEED}), which is naturally also 
local, see Ref.~\footnote{
Note that if we adopt the global-derivations approach, then our
additional unknown parameter $\beta\equiv(\partial B_x/\partial y)_o$
can be estimated as $\beta\approx B_x(0,L)/L\approx
(\delta_o/L)B_m/L\approx(V_R/V_{out})B_m/L$, the parameter 
$\upsilon\equiv(\partial V_y/\partial y)_o$ can be estimated
as $\upsilon\approx V_{out}/L$, and our equation~(\ref{SP_THIRD_MHD})
reduces to the Sweet-Parker equation~(\ref{FIRST_MHD}), as one
expects. 
}.

The reader of this paper could question why we develop and suggest 
a new theoretical approach to the problem of quasi-stationary
magnetic reconnection if we obtain the same results in the
Sweet-Parker reconnection case. The answer is that our approach,
based on local calculations, allows us to calculate the reconnection
rate in the case of anomalous resistivity and also provides an
additional important understanding of the reconnection problem. Our
derivations for anomalous reconnection are given in the next section
and we discuss our results in Sec.~\ref{CONCLUSIONS}. Note
that our local-equations approach to the reconnection problem and the
more conventional global-equations approach are the same from the
point of view of the underlying physics. Indeed, it is well known that
by using the Gauss-Ostrogradski theorem, most physical equations can
be written in two equivalent forms, in the form of local differential
equations and in the form of global integral equations.


\section{Magnetic reconnection with anomalous resistivity}
\label{NEW_MODEL}

In this section we study reconnection with anomalous resistivity
and derive a simple and accurate estimate of the reconnection
rate in the classical Sweet-Parker-Petschek two-and-a-half dimensional
reconnection layer shown in Fig.~\ref{FIG_RECONNECTION_LAYER}. We 
consider resistivity to be a given arbitrary function of the
z-component of the electric current and the two-dimensional
coordinates, $\eta=\eta(j_z,x,y)$, which has finite derivatives in $y$
up to the second order and in $x$ and $j_z$ up to the first 
order~\footnote{ 
We assume $\eta$ to be a function of $j_z$ instead of the total
current $j=(j_z^2+j_x^2+j_y^2)^{1/2}$. This is because the
reconnection process proceeds due to the z-component of the electric
field, see Eq.~(\ref{MHD_EQUATION}), and it is reasonable to assume
that the electrical conductivity in the z-direction can be reduced by
plasma instabilities due to large values of $j_z$.
}.

Let us list the assumptions that we make. 
First, we assume that the characteristic Lundquist number of the
problem is large, which (by our definition) is equivalent to the 
assumption that resistivity is negligible outside the
reconnection layer and the non-resistive MHD equations apply there. 
Second, we assume that the plasma flow is incompressible, 
${\rm div}\,{\bf V}=0$. In the limit of very high Lundquist numbers 
and slow reconnection rates the incompressibility condition is a very
good approximation in a reconnection layer even for compressible
plasmas~\cite{furth_1963}. 
Third, we assume that the reconnection process is quasi-stationary. 
This can only be the case if the reconnection rate is small, 
$\eta_oj_o/V_AB_m\approx V_R/V_A\ll1$ [$V_A\equiv B_m/\sqrt{\rho}$ and
see Eq.~(\ref{FIRST_MHD})], and there are no plasma instabilities in
the reconnection layer. Note that in our model the reconnection rate
can still be much faster than the Sweet-Parker rate. 
Fourth, we assume that the reconnection layer is thin,
$\delta_o/L'\ll1$. If plasma kinematic viscosity is small (in 
comparison with resistivity), then the plasma outflow velocity in the
downstream regions is equal to the Alfven velocity $V_A$, we
have plasma mass conservation condition $\delta_o/L'\approx V_R/V_A$,
and this assumption of a thin reconnection layer is fully equivalent
to the previous assumption of a small reconnection rate. However, if
plasma is very viscous, then the plasma outflow velocity is smaller
than $V_A$ and assumption $\delta_o/L'\ll1$ is stronger than
assumption $V_R/V_A\ll1$. Finally, note that we make no assumptions
about the values of the guide field $B_z$ and the plasma
viscosity. [However, below we will see that our assumption of a
quasi-stationary reconnection process in a thin current sheet layer
results in a necessary condition $\nu\ll\eta_o(V_A/V_R)$ for the
plasma kinematic viscosity $\nu$. Refer to Eq.~(\ref{NU}) for more
details.]

Now note that several equations that we derived in the previous
section for the case of the Sweet-Parker reconnection with constant
resistivity stay the same in the case of anomalous resistivity. 
Indeed, equation~(\ref{MHD_EQUATION}) stays valid when the
resistivity $\eta$ is not constant. Therefore,
equation~(\ref{FIRST_MHD}) also stays valid, except that $\eta_o$ now
is the value of resistivity at the reconnection layer central point~O
(see Fig.~\ref{FIG_RECONNECTION_LAYER}), 
i.e.~$\eta_o=\eta(j_z=j_o,x=0,y=0)$. Equations~(\ref{J_O})
and~(\ref{UPSILON_ESTIMATE}), which result from the Ampere's law and
the plasma incompressibility respectively, obviously stay valid 
too~\footnote{
Note that Eqs.~(\ref{J_O}) and~(\ref{UPSILON_ESTIMATE}) are exact for
the Harris model reconnection sheet~\cite{harris_1962}, which has 
$B_y=B_m\tanh(x/\delta_o)$, $B_x=0$, 
$j_z=(B_m/\delta_o)\cosh^{-2}(x/\delta_o)$, $\eta={\rm const}$ and
$V_x=-(\eta/\delta_o)\tanh(x/\delta_o)$.
}.
At the same time, in the general case of anomalous resistivity that we
consider in this section, we need to re-derive
equations~(\ref{SP_LOCAL_ENERGY}) and~(\ref{SP_THIRD_MHD}), which are
the equations of the plasma acceleration and spatial homogeneity
of the electric field z-component along the reconnection layer
(i.e.~along the y-axis). However, before we re-derive these two
equations, we would first like to derive the equation of magnetic
energy conservation. Using Eqs.~(\ref{FIRST_MHD}),~(\ref{J_O})
and~(\ref{UPSILON_ESTIMATE}), we immediately obtain 
\beq
\upsilon B_m^2=\epsilon\,\eta_oj_o^2,
\quad \epsilon\approx 1,
\label{B_ENERGY}
\eeq
where for the purpose of comparison of our theoretical results to
our numerical simulations we introduce a coefficient $\epsilon$, which
is of order of unity. Equation~(\ref{B_ENERGY}) is the equation for
magnetic energy conservation. The rate of the supply of magnetic
energy into the reconnection layer is equal to the rate of its
Ohmic dissipation inside the layer. 

Next we derive the equation of plasma acceleration along the
reconnection layer (i.e.~along the y-axis), taking into consideration
all forces acting on the plasma. The MHD equation for the y-component
of the plasma velocity $V_y$, assuming the quasi-stationarity of the
reconnection ($\partial/\partial t=0$) and plasma incompressibility 
(${\rm div}\,{\bf V}=0$), is~\cite{landau_1983}
\beq
\rho({\bf V}\cdot{\bf\nabla})V_y &=& -(\partial/\partial y)
\left(P+B_x^2/2+B_y^2/2\right)
\nonumber\\
&&{}+({\bf B}\cdot{\bf\nabla})B_y+\rho\nu{\bf\nabla}^2V_y,
\label{V_Y_EQUATION}
\eeq
where $\rho$ is the plasma density, $\nu$
is the plasma kinematic viscosity (assumed to be constant) and $P$ is
the sum of the plasma pressure and the guide field pressure
$B_z^2/2$. Taking the first order partial derivative
$\partial/\partial y$ of equation~(\ref{V_Y_EQUATION}) at the central
point~O, we obtain 
\beq
\rho\upsilon^2 = -(\partial^2 P/\partial y^2)_o+\beta j_o+
\rho\nu\left[{\bf\nabla}^2(\partial V_y/\partial y)\right]_o, 
\label{V_Y_EQUATION_2}
\eeq
where we use parameter $\beta\equiv(\partial B_x/\partial y)_o$ and
the Ampere's law 
$j_o=(\partial B_y/\partial x)_o-(\partial B_x/\partial y)_o$ at point
O, and we also use the formulas $V_x\equiv B_y\equiv0$ on the y-axis,
and $V_y=B_x=0$ at point O, which follow from the symmetry of the
problem with respect to the x- and y-axes. 
The pressure term $(\partial^2 P/\partial y^2)_o$ on the
right-hand-side of equation~(\ref{V_Y_EQUATION_2}) can be precisely
calculated in analogy with the Sweet-Parker derivation of the
pressure decline along the reconnection layer, which employs the 
force balance condition for the plasma across the reconnection layer
and leads to equation~(\ref{SP_ENERGY}). The viscosity term 
$\rho\nu\left[{\bf\nabla}^2(\partial V_y/\partial y)\right]_o$ in
equation~(\ref{V_Y_EQUATION_2}) can be calculated approximately by
using estimates for the $V_y$ velocity derivatives.
In Appendix~\ref{APPENDIX_A} we carry out these calculations and
show that the pressure and viscosity terms are equal to
\beq
(\partial^2 P/\partial y^2)_o &=&
B_m(\partial^2 B_y/\partial y^2)_m
\nonumber\\
&&{}+{\rm o}\{\rho\upsilon^2,\beta j_o,\rho\nu\upsilon/\delta_o^2\},
\qquad
\label{PRESSURE_TERM}
\\
\rho\nu\left[{\bf\nabla}^2(\partial V_y/\partial y)\right]_o
&\approx& -\rho\nu\upsilon/\delta_o^2,
\label{VISCOSITY_TERM}
\eeq
where $B_m(\partial^2 B_y/\partial y^2)_m$ is calculated just
outside the reconnection layer at point M (see
Fig.~\ref{FIG_RECONNECTION_LAYER}), and expression
${\rm o}\{\rho\upsilon^2,\beta j_o,\rho\nu\upsilon/\delta_o^2\}$
denotes terms that are small compared to either $\rho\upsilon^2$ or
$\beta j_o$ or $\rho\nu\upsilon/\delta_o^2$ in the limit of
a slow reconnection rate in a thin reconnection 
layer, see Ref.~\footnote{
Note that the 
${\rm o}\{\rho\upsilon^2,\beta j_o,\rho\nu\upsilon/\delta_o^2\}$ terms
can still be much larger than the $B_m(\partial^2 B_y/\partial y^2)_m$
term in equation~(\ref{PRESSURE_TERM}), in which case the pressure
term $(\partial^2 P/\partial y^2)_o$ is unimportant and negligible 
in equation~(\ref{V_Y_EQUATION_2}). This happens when reconnection
with anomalous resistivity is much faster than Sweet-Parker
reconnection. 
}.
Substituting equations~(\ref{PRESSURE_TERM})
and~(\ref{VISCOSITY_TERM}) into equation~(\ref{V_Y_EQUATION_2}) and
using equation~(\ref{J_O}) for $\delta_o$, we obtain 
\beq
\rho\upsilon^2&\approx&-B_m(\partial^2 B_y/\partial y^2)_m
+\beta j_o-\rho\nu\upsilon j_o^2/B_m^2,
\qquad
\label{SECOND_MHD}\\
&&\quad\mbox{where~}
\beta\equiv(\partial B_x/\partial y)_o\mbox{~~at point O}.
\nonumber
\eeq
Note that this equation is exact if plasma viscosity can be 
neglected ($\nu=0$).

Now we use the condition of spatial homogeneity of the electric field
z-component along the reconnection layer, i.e.~along the 
y-axis. We take the second order partial derivatives 
$\partial^2/\partial y^2$ of the left- and right-hand-sides of 
equation~(\ref{MHD_EQUATION}) at the central point O (note that the
first order partial derivatives are identically zero). Taking into
account the symmetry of the problem, so that $V_x\equiv B_y\equiv0$ on
the y-axis, and $V_y=B_x=\partial j_z/\partial y=0$ at point O, we
obtain 
\beq
&&-\left[\eta_o+j_o(\partial\eta/\partial j_z)_o\right]
(\partial^2 j_z/\partial y^2)_o-j_o(\partial^2\eta/\partial y^2)_o
\qquad
\nonumber\\
&&\qquad\qquad
=2(\partial V_y/\partial y)_o(\partial B_x/\partial y)_o= 
2\upsilon\beta,
\label{THIRD_MHD}
\eeq
where we use formulas $\upsilon\equiv(\partial V_y/\partial y)_o$ and
$\beta\equiv(\partial B_x/\partial y)_o$. 
Finally, we need to estimate the $(\partial^2 j_z/\partial y^2)_o$
term, which enters the left-hand-side of equation~(\ref{THIRD_MHD}).
This estimation can be done by taking the second order partial
derivative $\partial^2/\partial y^2$ of equation~(\ref{J_O}), while
keeping $\delta_o$ constant because the partial derivative in $y$ is
to be taken at a constant value $x={\rm const}=\delta_o$. In
Appendix~\ref{APPENDIX_B} we give the detailed derivations and find
that the y-scale of the current $j_z$ is about the same as the y-scale
of the outside magnetic field,  
i.e.~$j_o^{-1}(\partial^2 j_z/\partial y^2)_o\approx 
B_m^{-1}(\partial^2 B_y/\partial y^2)_m$. However, for the purpose of
comparison of our theoretical results to numerical simulations in
Sec.~\ref{SIMULATIONS}, we find it convenient to write 
\beq
j_o^{-1}(\partial^2 j_z/\partial y^2)_o=
\gamma B_m^{-1}(\partial^2 B_y/\partial y^2)_m,
\quad\gamma\approx 1,
\label{J_YY}
\eeq
where we introduce the coefficient $\gamma$, which is of order unity.

Let us take the dimensionless coefficients $\epsilon$ and $\gamma$,
which enter equations~(\ref{B_ENERGY}) and~(\ref{J_YY}), and are of
order unity, to be exactly unity, $\epsilon=1$ and $\gamma=1$. 
Now we have all the equations necessary to determine all unknown 
physical parameters. In particular, using
Eqs.~(\ref{B_ENERGY}),~(\ref{SECOND_MHD}),~(\ref{THIRD_MHD})
and~(\ref{J_YY}), we easily obtain the following approximate algebraic
equation for the z-current $j_o$ at the reconnection layer central
point O: 
\beq
&&3+\frac{j_o(\partial\eta/\partial j_z)_o}{\eta_o}
+\frac{B_m(\partial^2\eta/\partial y^2)_o}
{\eta_o(\partial^2 B_y/\partial y^2)_m}
\nonumber\\
&&\qquad\approx-\left(1+\frac{\nu}{\eta_o}\right)
\frac{\eta_o^2j_o^4}{V_A^2B_m^4}
\frac{2B_m}{(\partial^2 B_y/\partial y^2)_m},
\qquad
\label{RATE}
\eeq
where the Alfven velocity $V_A$ is defined as
$V_A\equiv B_m/\sqrt{\rho}$ and $\eta_o=\eta(j_z=j_o,x=0,y=0)$ is the
resistivity at point O. Given the resistivity function
$\eta=\eta(j_z,x,y)$, as well as the magnetic 
field $B_m$ and its second order derivative 
$(\partial^2 B_y/\partial y^2)_m$ outside the reconnection layer, we
can solve equation~(\ref{RATE}) for the current $j_o$ and find the
reconnection rate, which is the rate of destruction of magnetic flux
at point O, equal to $-(\partial A_z/\partial t)_o$.  
Using Eq.~(\ref{MHD_EQUATION}), we find that the reconnection rate is
equal to $\eta_oj_o=E_z$. Note that for the classical reconnection
layer that we consider (see Fig.~\ref{FIG_RECONNECTION_LAYER}) the
right-hand-side of equation~(\ref{RATE}) is positive because 
$2B_m/(\partial^2 B_y/\partial y^2)_m\approx-L^2<0$, where $L$
is the global scale of the magnetic field outside the reconnection
layer. Once the current $j_o$ is calculated by means of
equation~(\ref{RATE}), we can easily calculate all other reconnection
parameters, using
Eqs.~(\ref{FIRST_MHD}),~(\ref{J_O}),~(\ref{UPSILON_ESTIMATE})
and~(\ref{SECOND_MHD}),
\beq
V_R &\approx& \eta_oj_o/B_m\ll V_A,
\label{V_R}
\\
\upsilon &\approx& \eta_oj_o^2/B_m^2,
\label{UPSILON}
\\
\beta &\approx& j_o\left[(1+\nu/\eta_o)(\eta_o j_o/V_AB_m)^2\right.
\nonumber\\
&&\quad\;\left.
+(B_m/j_o^2)(\partial^2 B_y/\partial y^2)_m\right]\ll j_o, 
\label{BETA}
\\
\delta_o &\approx& B_m/j_o \approx V_R/\upsilon.
\label{DELTA}
\eeq
Equations~(\ref{RATE})-(\ref{DELTA}) are the most general result for
magnetic reconnection that we obtain in this paper. Restoring
coefficients $\epsilon\approx1$ and $\gamma\approx1$,
equation~(\ref{RATE}) becomes
\beq
&&(\gamma+2\epsilon)+
\gamma\frac{j_o(\partial\eta/\partial j_z)_o}{\eta_o}
+\frac{B_m(\partial^2\eta/\partial y^2)_o}
{\eta_o(\partial^2 B_y/\partial y^2)_m}
\nonumber\\
&&\qquad=-\epsilon^2\left(\epsilon+\frac{\nu}{\eta_o}\right)
\frac{\eta_o^2j_o^4}{V_A^2B_m^4}
\frac{2B_m}{(\partial^2 B_y/\partial y^2)_m}.
\qquad
\label{RATE_GENERAL}
\eeq

Hereafter we will consider the natural case when 
$(\partial\eta/\partial j_z)_o\ge0$ and 
$(\partial^2\eta/\partial y^2)_o\le0$ because plasma conductivity
decreases as the current increases and we are interested in fast
anomalous reconnection (i.e.~faster than the Sweet-Parker
reconnection). In this case the first, second and third terms on the
left-hand-side of equation~(\ref{RATE}) are all positive. It is easy
to see that the first term is related to Sweet-Parker reconnection
with constant resistivity equal to $\eta_o$, the second term is
related to fast reconnection associated with the dependence of
anomalous resistivity on the current, and the third term is related to
fast reconnection associated with an {\it ad hoc} localization of
resistivity in space (see the next section for details). Also note
that if the plasma kinematic viscosity $\nu$ is larger than the
resistivity $\eta_o$, then, according to equation~(\ref{RATE}), the
current $j_o$ and reconnection rate $\eta_oj_o$ become smaller as
$\nu$ grows, i.e.~the reconnection slows down for viscous plasmas as
one expects.

We postpone the analysis of equations~(\ref{RATE})-(\ref{DELTA}) until
Sec.~\ref{CONCLUSIONS}. Let us now make an estimate of the
half-length of the reconnection layer $L'$ (see
Fig.~\ref{FIG_RECONNECTION_LAYER}). Note that $L'$ is not needed for
the calculation of the reconnection rate $\eta_oj_o$ by means of
equation~(\ref{RATE}). Nevertheless, we are still interested in a
rough estimate of $L'$, in particular, because we need to check our
assumption that the reconnection layer is thin, i.e.~that the
condition $\delta_o\ll L'$ is satisfied. It is clear that $L'$ can not
be much larger than the global scale of the magnetic field outside the 
layer $L$. Therefore we have the condition $L'\lesssim L$. However,
$L'$ can be much smaller than $L$, in which case the reconnection
layer has a pair of the Petschek switch-off MHD shocks attached to
each end of the layer in the downstream
regions~\cite{petschek_1964,kulsrud_2001,vasyliunas_1975}, as shown in
Fig.~\ref{FIG_RECONNECTION_LAYER}. In this case $L'$ should be
calculated as the y-coordinate of the point on the y-axis at which the
perpendicular field $B_x$ is strong enough to support the
shocks~\cite{kulsrud_2001}. Following Kulsrud~\cite{kulsrud_2001}, we
use the jump condition on the Petschek switch-off shocks to 
obtain~\footnote{
The jump condition~(\ref{SCHOCK_BALANCE}) was used by
Kulsrud~\cite{kulsrud_2001} for the case of a viscosity-free
plasma. It can be shown from the full non-ideal MHD equations that
this condition is unchanged in the case of a viscous plasma.
}
\beq
V_R\approx B_x(0,y=L')/\sqrt{\rho}\approx\beta L'/\sqrt{\rho}=
\beta L'V_A/B_m,\quad
\label{SCHOCK_BALANCE}
\eeq
where we use the first-order Taylor expansion for an estimate 
$B_x(0,y=L')\approx(\partial B_x/\partial y)_oL'=\beta L'$ and, as
before, $V_A\equiv B_m/\sqrt{\rho}$. Now, using 
Eqs.~(\ref{V_R}),~(\ref{BETA}) and~(\ref{SCHOCK_BALANCE}), we
can easily find $L'$. Before we write the explicit formula for $L'$,
note that the absolute value of the 
$(B_m/j_o^2)(\partial^2 B_y/\partial y^2)_m$ term in 
equation~(\ref{BETA}) is equal or smaller than the 
$(1+2\nu/\eta_o)(\eta_oj_o/V_AB_m)^2$ term. This follows from the fact
that the left-hand-side of equation~(\ref{RATE}) is equal or greater
than unity [see our comments in the paragraph that follows
Eq.~(\ref{RATE_GENERAL})]. Therefore, the 
$(B_m/j_o^2)(\partial^2 B_y/\partial y^2)_m$ term can be omitted in 
equation~(\ref{BETA}) for the purpose of estimating $L'$. As a
result, we obtain the following rough estimates for the reconnection
layer half-length $L'$ and the velocity of plasma outflow in the
downstream regions $V_{out}$:
\beq
L' &\approx& (V_AB_m^2/\eta_oj_o^2)(1+\nu/\eta_o)^{-1}
\nonumber\\
&\approx&
(V_A/\upsilon)(1+\nu/\eta_o)^{-1},
\qquad \delta_o \ll L'\lesssim L,
\qquad
\label{L'}
\\
V_{out} &\approx& \upsilon L'\approx V_A(1+\nu/\eta_o)^{-1}
\le V_A,
\label{V_OUT}
\eeq
where we use equations~(\ref{V_R}),~(\ref{UPSILON}),~(\ref{BETA})
and~(\ref{SCHOCK_BALANCE}). Note that the condition 
$L'\lesssim L$ is always satisfied because 
$2B_m/(\partial^2 B_y/\partial y^2)_m\approx-L^2$, and the left- and
right-hand-sides of equation~(\ref{RATE}) are equal or greater than
unity. However, the condition that the reconnection layer is thin,
$\delta_o\ll L'$, is satisfied only if plasma viscosity is not too
large, 
\beq
\nu\ll \eta_o(V_AB_m/\eta_oj_o)\approx\eta_o(V_A/V_R),
\label{NU}
\eeq
where, to derive this formula, we use
Eqs.~(\ref{V_R}),~(\ref{DELTA}) and~(\ref{L'}). In other words,
to be able to form a thin reconnection layer, the plasma should not be
too viscous. Note that in the case of constant resistivity and large
viscosity, $\eta_o={\rm const}$ and $\nu\gg\eta_o$, the reconnection
velocity is
$V_R/V_A\approx(\eta_o/V_AL)^{1/2}(\eta_o/\nu)^{1/4}$
(see~\cite{park_1984}) and condition~(\ref{NU}) reduces to
$\nu\ll\eta_o(V_AL/\eta_o)^{2/3}=V_AL(\eta_o/V_AL)^{1/3}$. Finally, if
the plasma viscosity is small in comparison to the resistivity, 
$\nu\ll\eta_o$, then from equation~(\ref{V_OUT}) we immediately find
an important and well-known result that in this case the velocity of
the plasma outflowing in the downstream regions at the ends of the
reconnection layer is approximately equal to the Alfven velocity,  
$V_{out}\approx V_A$ (see Fig.~\ref{FIG_RECONNECTION_LAYER}).

At the end of this section we would like to discuss several
assumptions that we used in our derivations. First, the solution of
equation~(\ref{RATE}) is valid only if it gives $\eta_oj_o\ll V_AB_m$,
which is our assumption of a slow quasi-stationary reconnection. 
Because of equation~(\ref{V_R}), condition $\eta_oj_o\ll V_AB_m$ is
equivalent to $V_R\ll V_A$, i.e.~the reconnection velocity, which is
the velocity of the incoming plasma, must be small in comparison to
the Alfven velocity in the upstream region. Second, the coefficient 
$\beta\equiv(\partial B_x/\partial y)_o$, given by
Eq.~(\ref{BETA}), must be much smaller than the current 
$j_o=(\partial B_y/\partial x)_o-\beta\approx
(\partial B_y/\partial x)_o$ because the reconnection layer
is assumed to be thin. It is easy to see that this condition is
satisfied. Indeed, the first term in the brackets $[...]$ in
equation~(\ref{BETA}) is much smaller than unity because of the upper
limit for plasma viscosity given by equation~(\ref{NU}). The second
term in the brackets in equation~(\ref{BETA}) is also much smaller
than unity because $j_o$ is much larger than the electric current
outside the reconnection layer due to our assumption of a large
characteristic Lundquist number of the system.


\section{Special cases of magnetic reconnection}
\label{RECONNECTION_LIMITS}

In this section we focus on three special cases for the reconnection
rate, which arise when one of the three terms on the left-hand-side of
equation~(\ref{RATE}) dominates over the other two.
We consider the classical Sweet-Parker-Petschek reconnection layer
shown in Fig.~\ref{FIG_RECONNECTION_LAYER} and define the global
scale of the magnetic field outside the reconnection layer as
$L\equiv\sqrt{-2B_m/(\partial^2 B_y/\partial y^2)_m}$ 
(see~\footnote{
Following Kulsrud~\cite{kulsrud_2001}, for the definition of the
global magnetic field scale $L$ we use formula
$B_y(\delta_o,y)=B_m(1-y^2/L^2)$ for the field y-component along the
interval M${\tilde{\rm M}}$ that is outside the reconnection layer as
shown in Fig.~\ref{FIG_RECONNECTION_LAYER}.
}).
In addition, for the purpose of clarity, in this section we focus only
on resistivity effects and neglect plasma viscosity, assuming
that $\nu\ll\eta_o$.

\subsection{Sweet-Parker reconnection, \boldmath$\eta={\rm const}$}
\label{SWEET_PARKER_RECONNECTION}

\begin{figure}[t]
\vspace{6.0truecm}
\includegraphics{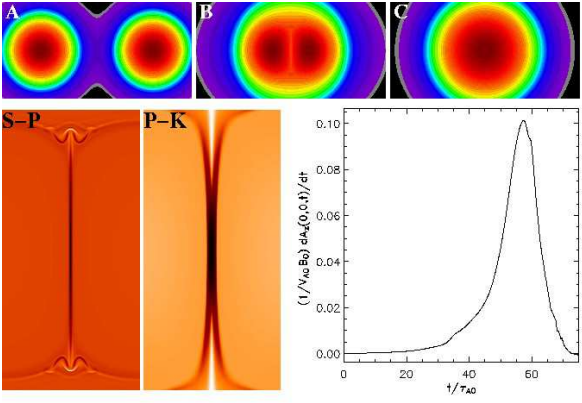}
\caption{(Color online) Three top plots: merging of two cylindrical
magnetic flux tubes by reconnection ($A_z$ is plotted). Two
bottom-left plots: the configuration of the reconnection layer for the
Sweet-Parker (\mbox{``S-P''}) and Petschek-Kulsrud (\mbox{``P-K\-''})
cases ($j_z$ is plotted). The bottom-right plot: the reconnection rate
$dA_z/dt=\eta_oj_o$ at the layer central point O, normalized by
$V_{A0}B_0$. 
} 
\label{FIG_ROSETTE}
\end{figure}

First, consider the case when resistivity is constant,
$\eta(j_z,x,y)=\eta_o={\rm const}$. In this case only the first term
on the left-hand-side of equation~(\ref{RATE}) is nonzero and
equations~(\ref{RATE})-(\ref{DELTA}) and~(\ref{L'}) reduce to 
\beq
\begin{array}{lcl}
3^{1/2}\approx\eta_o j_o^2L/V_AB_m^2 
&\;\Rightarrow\;& j_o\approx(B_m/L)S_o^{1/2},
\\
S_o\equiv V_AL/\eta_o\gg1, 
&& V_R\approx V_AS_o^{-1/2},
\\
\upsilon\approx V_A/L,
&& \beta\approx(B_m/L)S_o^{-1/2}, 
\\
\delta_o\approx LS_o^{-1/2},
&& L'\approx L, 
\end{array}
\label{SP_RECONNECTION}
\eeq
where we set the plasma kinematic viscosity to zero ($\nu=0$), use our
definition of the global field scale 
$L\equiv\sqrt{-2B_m/(\partial^2 B_y/\partial y^2)_m}$, introduce the
Lundquist number $S_o\equiv V_AL/\eta_o$ and assume for our estimates
that $3^{1/4}\approx1$. The above equations are the Sweet-Parker
reconnection equations with constant resistivity equal to
$\eta_o$. Thus, we find that if resistivity is constant,
then the reconnection must be Sweet-Parker and not
Petschek~\cite{kulsrud_2001,kulsrud_2005}. We discuss this important
result in Sec.~\ref{CONCLUSIONS}. Note that in this section,
contrary to Sec.~\ref{SWEET_PARKER_MODEL}, we do not assume the
Sweet-Parker geometrical configuration for the reconnection layer, but
derive it together with the reconnection rate from our general
equations of the previous section. A typical configuration of the
reconnection layer in the case of Sweet-Parker reconnection is shown
on the left-bottom plot in Fig.~\ref{FIG_ROSETTE}. This plot is
marked by letters ``S-P''.

\subsection{Petschek-Kulsrud reconnection,
\boldmath$\eta=\eta(j_z)$}
\label{PETSCHEK_KULSRUD_RECONNECTION}

Now let us consider the case when resistivity is anomalous and is a
monotonically increasing function of the electric current only,
$\eta=\eta(j_z)$. Let us further assume that this dependence of
resistivity on the current is very strong, so that
$(j_o/\eta_o)(d\eta/dj_z)_o\gg1$. In this case the
second term on the left-hand-side of equation~(\ref{RATE}) is
dominant, and equations~(\ref{RATE}) and~(\ref{L'}) reduce to 
\beq 
(d\eta/d j_z)_o &\approx& \eta_o^3j_o^3L^2/V_A^2B_m^4
\;\;\Rightarrow{}
\nonumber\\
\frac{V_R}{V_A} &\approx& \frac{\eta_oj_o}{V_AB_m}\approx
\frac{\delta_o}{L'}\approx\left[\frac{B_m}{V_AL^2}
\left(\frac{d\eta}{dj_z}\right)_{\!\!o}\,\right]^{1/3}
\!\!\!\!\!,\qquad
\label{PK_RECONNECTION_RATE}
\\
L'&\approx& \frac{V_AB_m^2}{\eta_oj_o^2}
\equiv L\frac{(\eta_oj_o/V_AB_m)^{-3/2}}
{(V_Aj_oL^2/\eta_oB_m)^{1/2}}
\nonumber\\
&\approx&
L\left[\frac{j_o}{\eta_o}
\left(\frac{d\eta}{dj_z}\right)_{\!\!o}\,\right]^{-1/2}
\ll L.
\label{PK_RECONNECTION_L'}
\eeq
Here again we set $\nu=0$, and we use the formula
$L^2\equiv-2B_m/(\partial^2 B_y/\partial y^2)_m$ and
Eqs.~(\ref{V_R}) and~(\ref{DELTA}). From
equation~(\ref{PK_RECONNECTION_L'}) we see that the half-length of the
reconnection layer $L'$ is much less than the global field scale $L$
in the case of a strong dependence of resistivity on the current 
($(j_o/\eta_o)(d\eta/dj_z)_o\gg1$). This means that in this case the
geometry of the reconnection layer is Petschek with a pair of shocks
attached to each end of the layer in the downstream regions, as shown
in Fig.~\ref{FIG_RECONNECTION_LAYER}. 
Equation~(\ref{PK_RECONNECTION_RATE}) for the reconnection rate was
first analytically derived by Kulsrud~\cite{kulsrud_2001} for a
special case when $d\eta/dj_z\equiv{\rm const}=\eta_\star/j_c$. 
In the Petschek-Kulsrud reconnection case the reconnection rate, given
by Eq.~(\ref{PK_RECONNECTION_RATE}), can be considerably faster than
the Sweet-Parker reconnection rate. This fast reconnection has been
previously observed in many numerical simulations done with anomalous
resistivity $\eta=\eta(j_z)$
(e.g.~see~\cite{hayashi_1978,sato_1979,breslau_2003}). The 
typical configuration of the reconnection layer in the case of
Petschek-Kulsrud reconnection is shown in the left-bottom plot in
Fig.~\ref{FIG_ROSETTE}, this plot is marked by letters ``P-K''.

\subsection{Spatially localized reconnection,
\boldmath$\eta=\eta(x,y)$}
\label{SPATIALLY_LOCALIZED_RECONNECTION}

Finally let us consider the special case when resistivity is given by
$\eta=\eta(x,y)$ and it is spatially localized around the central 
point O of the reconnection layer (see
Fig.~\ref{FIG_RECONNECTION_LAYER}), so that 
$-2\eta_o/(\partial^2\eta/\partial y^2)_o\equiv l_\eta^2\ll 
L^2\equiv -2B_m/(\partial^2 B_y/\partial y^2)_m$. In other words, we
assume that resistivity is anomalous and is localized on scale
$l_\eta$ that is much smaller than the global field scale $L$. In this
case the third term on the left-hand-side of equation~(\ref{RATE}) is
dominant, and equations~(\ref{RATE})-(\ref{DELTA}) and~(\ref{L'})
reduce to 
\beq 
\begin{array}{lcl}
L/l_\eta\approx\eta_o j_o^2L/V_AB_m^2 
&\;\Rightarrow\;& j_o\approx(B_m/l_\eta)S_l^{1/2},
\\
S_l\equiv V_Al_\eta/\eta_o\gg1,
&& V_R\approx V_AS_l^{-1/2},
\\
\upsilon\approx V_A/l_\eta,
&& \beta\approx(B_m/l_\eta)S_l^{-1/2}, 
\\
\delta_o\approx l_\eta S_l^{-1/2},
&& L'\approx l_\eta\ll L, 
\end{array}
\label{LOCALIZED_RECONNECTION}
\eeq
where we again set $\nu=0$. The above equations are the
same as the Sweet-Parker equations~(\ref{SP_RECONNECTION}) with the
global field scale $L$ replaced by the resistivity scale 
$l_\eta\ll L$. Note that, when resistivity is
localized, the reconnection rate becomes faster than the Sweet-Parker
rate by a factor $\sqrt{L/l_\eta}\gg 1$ and the geometry of the
reconnection layer is Petschek with a pair of shocks attached to each
end of the layer (see Fig.~\ref{FIG_RECONNECTION_LAYER}). These
results are in agreement with many previous numerical simulations of
reconnection with spatially localized
resistivity~\cite{ugai_1977,tsuda_1977,scholer_1989,biskamp_2001}.


\section{Numerical simulations of magnetic reconnection}
\label{SIMULATIONS}

In this section we present the results of our numerical simulations
of unforced reconnection of two cylindrical magnetic flux tubes. 
These simulations are not intended as a check or a proof of our
theoretical results for magnetic reconnection. Our 
equations~(\ref{RATE})-(\ref{DELTA}) have been derived
analytically and are very general. A comprehensive testing of them
would require extensive computational work, which is beyond the
scope of this paper. Instead, we present our simulations as a
demonstration of our reconnection model predictions. 

Following Kulsrud~\cite{kulsrud_2001}, we assume that plasma
resistivity is given by the following piecewise linear
function of the z-component of the electric current:
\beq
\eta(j_z)=\eta_s+\eta_\star\max\,\{0,j_z-j_c\}/j_c,
\label{LINEAR_RESISTIVITY}
\eeq
where $\eta_s$ is the Spitzer resistivity, which is assumed to be 
very small, $\eta_\star$ is the anomalous resistivity parameter and
$j_c$ is the critical current parameter. The Kulsrud model's
prediction for the reconnection rate in the case
$(j_o/\eta_o)(d\eta/dj_z)_o=j_o\eta_\star/\eta_oj_c\gg1$, which is
given by equation~(\ref{PK_RECONNECTION_RATE}), has already been 
checked and confirmed numerically by Breslau and
Jardin~\cite{breslau_2003}. Here we simulate reconnection with
anomalous resistivity given by equation~(\ref{LINEAR_RESISTIVITY}) for
a different computational setup, a higher Lundquist number and a wider
range of parameters $\eta_\star$ and $j_c$ (without the restriction
$j_o\eta_\star/\eta_oj_c\gg1$). Our intent is to see how our general 
formula~(\ref{RATE}) for the reconnection rate works in this case.

We consider an unforced reconnection of two cylindrical magnetic flux
tubes with the initial z-component of the magnetic field vector
potential equal to 
\beq
A_z(x,y)&=&A_0\left[\exp{\left(-r_+^2/2R_0^2\right)}+
\exp{\left(-r_-^2/2R_0^2\right)}\right],
\nonumber\\
A_0&=&B_0R_0\sqrt{e},
\quad r_\pm^2=(x\mp d)^2+y^2,
\label{TUBES}
\eeq
see the top-left plot in Fig.~\ref{FIG_ROSETTE}.
This convenient computational setup was suggested to us by Mikic
and Vainshtein~\cite{mikic_2003}. We choose the
parameters in equation~(\ref{TUBES}) as $R_0=1$ (the global scale of
the field is unity), $d=2.7162$ ($5$\% of initially reconnected flux)
and $B_0=1$ (the maximal initial field is unity). Thus,
$A_0=\sqrt{e}$. In addition, for convenience we choose the plasma
density $\rho_0=1$, so that the typical Alfven velocity and time are
unity, $V_{A0}\equiv B_0/\sqrt{\rho_0}=1$ and 
$\tau_{A0}\equiv R_0/V_{A0}=1$. The guide field is chosen to be zero,
$B_z=0$, and the initial plasma velocities are zero. The initial gas
pressure $P$ is chosen in such way that each of the two cylindrical
magnetic flux tubes~(\ref{TUBES}) would initially be in complete
equilibrium, $P+B_x^2/2+B_y^2/2={\rm const}$, if there were no
magnetic forces from the other tube. The plasma kinematic viscosity is
chosen to be equal to the Spitzer resistivity, $\nu=\eta_s$. The
boundary conditions are placed at $x,y=\pm25$, which are virtually at
infinity (the magnetic vector potential~(\ref{TUBES}) drops to less
than $10^{-100}$ at the boundaries). Because of the symmetry of the
problem, in the case of a quasi-stationary reconnection considered
here, it is enough to run simulations only in the upper-right
quadrangle of the full computational box.  
We use the FLASH code for our simulations. This is a compressible
adaptive-mesh-refinement (AMR) code written and supported at the ASC
Center of the University of Chicago. (For a comprehensive description
of the FLASH code see~\cite{fryxell_2000,calder_2002}). The MHD module
of the code uses central finite differences to properly resolve all
resistive and viscous scales. Comparing numerical results obtained
by simulations done with a compressible code to our approximate
theoretical formulas derived for incompressible fluids is fine in a
case of a very high Lundquist number. This is because in this case the
incompressibility condition is a very good approximation in a
reconnection layer even for compressible plasmas~\cite{furth_1963}.
Indeed, in our simulations the plasma density varies by no more than
15\% inside the reconnection layer. The biggest advantage of the FLASH
code for our purposes is that it is an already existing, well tested
code with the AMR feature, which allows us to place the boundary
conditions at infinity. The size of the smallest elementary grid cell
in our two-dimensional simulations was chosen typically to vary from
$25/2^{15}=0.0007629$ to $25/2^{13}=0.003052$,
which was sufficient to resolve the resistive reconnection layer.

The two cylindrical magnetic flux tubes, initially set up according to
equation~(\ref{TUBES}), attract each other (in a similar way as two
wires with colliniar currents do). As a result, as time goes on, the
tubes move toward each other, form a thin reconnection layer along the
y-axis and eventually completely merge together by reconnection. This
merging process is displayed in the three top plots in
Fig.~\ref{FIG_ROSETTE}, which show the field vector potential
$A_z$ in a central region of the full computational box. The two
bottom-left plots in Fig.~\ref{FIG_ROSETTE} show the electric
current $j_z$ in a central region that includes the reconnection
layer. These plots clearly show the reconnection layer configuration
which is formed during the reconnection process in the 
cases of the Sweet-Parker and Petschek-Kulsrud reconnection (refer to
Secs.~\ref{SWEET_PARKER_RECONNECTION}
and~\ref{PETSCHEK_KULSRUD_RECONNECTION}). The bottom-right plot in
Fig.~\ref{FIG_ROSETTE} demonstrates the functional
dependence on time typical of the normalized reconnection rate
$(1/V_{A0}B_0)dA_z/dt=\eta_oj_o/V_{A0}B_0$ at the reconnection
layer central point O. (This point is shown in 
Fig.~\ref{FIG_RECONNECTION_LAYER}). Next we compare the maximal
(peak) reconnection rate observed in the numerical simulations with
the theoretical rate predicted by our reconnection model.

\begin{figure}[t]
\vspace{10.5truecm}
\includegraphics{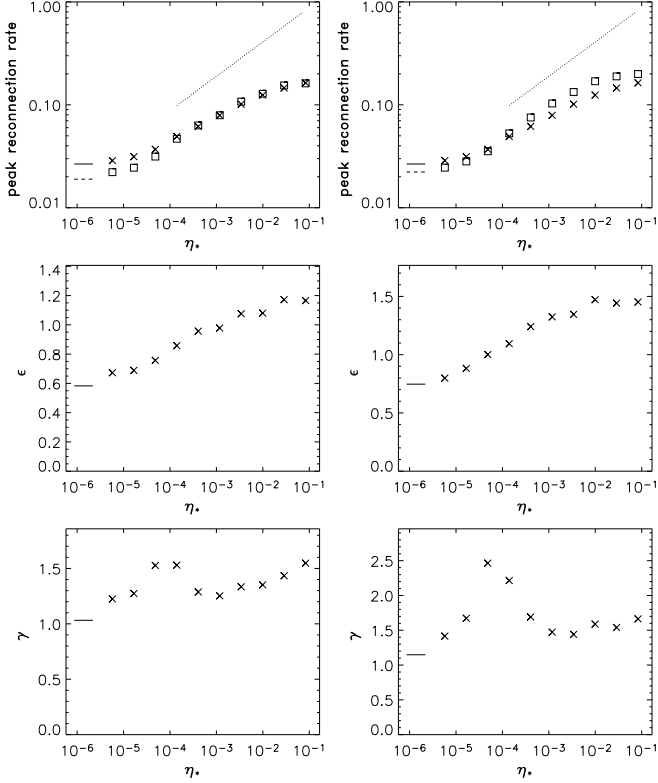}
\caption{Three left plots: the reconnection rate (left-top),
coefficient $\epsilon$ (left-central) and coefficient $\gamma$
(left-bottom) as functions of $\eta_\star$ at fixed $\eta_s=0.0002$ 
and $j_c=17.100$. The crosses and boxes show the results of our
simulations and theory respectively. The solid/dashed horizontal lines
correspond to $\eta_\star=0$ and simulations/theory. The inclined
dotted line given in the reconnection rate plot shows the
$\propto\eta_\star^{1/3}$ scaling. Point M (see
Fig.~\ref{FIG_RECONNECTION_LAYER}) is chosen to satisfy 
$(\eta j_z)_m=(1/3)\eta_oj_o$. Three right plots: the same as the
three left plots except point M is chosen to satisfy 
$(\eta j_z)_m=(1/10)\eta_oj_o$.  
} 
\label{FIG_RATES_1_3}
\end{figure}

\begin{figure}[t]
\vspace{10.5truecm}
\includegraphics{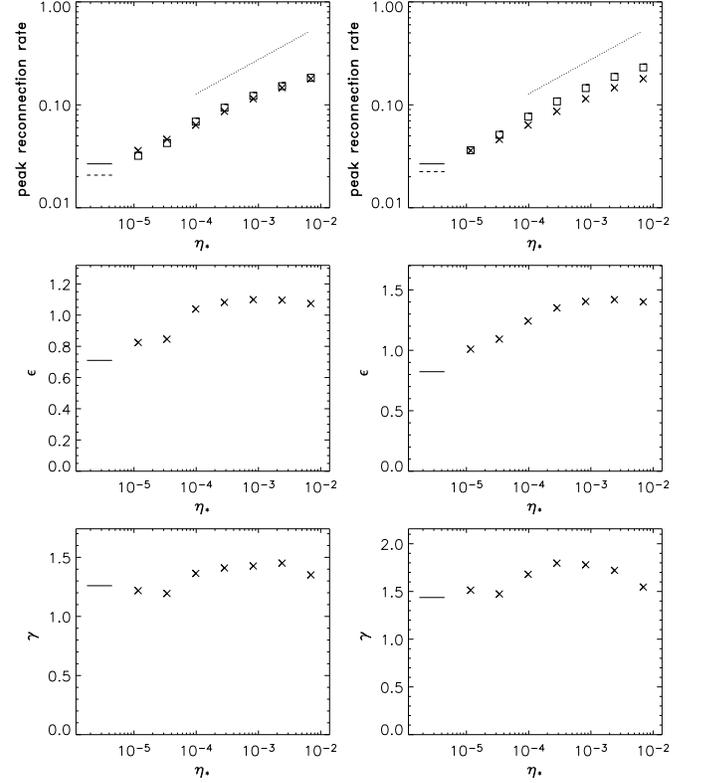}
\caption{These plots are the same as those in
Fig.~\ref{FIG_RATES_1_3}, except here $j_c=4.135$. 
} 
\label{FIG_RATES_2_3}
\end{figure}

When resistivity is given by
equation~(\ref{LINEAR_RESISTIVITY}) and $\nu=\eta_s$, our
theoretical formula~(\ref{RATE}) for the reconnection rate reduces to
\beq
3+\frac{\eta_\star}{j_c}\frac{j_o}{\eta_o}
\approx(1+\eta_s/\eta_o)\frac{\eta_o^2j_o^4R_0^2}{V_{A0}^2B_0^4}
\;\frac{B_0^6}{B_m^6}\;\frac{L^2}{R_0^2}\;\frac{\rho}{\rho_0},
\label{LINEAR_MODEL_RATE}
\eeq
where, as explained above, in our numerical simulations we choose
$B_0=1$, $R_0=1$, $V_{A0}=1$ and $\rho_0=1$, and by definition the
global field scale is
$L\equiv\sqrt{-2B_m/(\partial^2 B_y/\partial y^2)_m}$.
We also assume that $j_o>j_c$, which is the case that we consider in
our numerical simulations. As we can see from
equation~(\ref{LINEAR_MODEL_RATE}), the theoretical reconnection rate
$\eta_oj_o$ is rather sensitive to $B_m$ and $L$, which are the
strength and scale of the magnetic field at point M outside the
reconnection layer (see Fig.~\ref{FIG_RECONNECTION_LAYER}). 
Therefore, in order to compare the theoretical results with the
results of our simulations, we need to accurately calculate the $B_m$
and $L$ observed in the simulations. As a result, the choice of the
exact position of point M, at which $B_m$ and $L$ are calculated, is
important. First, in our simulations we choose the point M to be the
point on the x-axis at which the observed resistivity term $\eta j_z$
is three times smaller than that observed at the central point O, 
i.e.~$(\eta j_z)_m=(1/3)\eta_oj_o$. The three plots on the left in
Fig.~\ref{FIG_RATES_1_3} demonstrate our results for this choice.
The top plot on the left shows the reconnection rate. The crosses
are a log-log plot of the maximal (peak) reconnection rate observed
in our simulations as a function of parameter $\eta_\star$ for 
fixed $\eta_s=0.0002$ and $j_c=17.100$. The boxes are the theoretical
reconnection rate, which is given by
equation~(\ref{LINEAR_MODEL_RATE}) with the appropriate values of 
$B_m$, $L$ and density $\rho$ observed in the simulations. The solid
horizontal line (simulations) and the dashed horizontal line (theory) 
correspond to the $\eta_\star=0$ case. The inclined dotted line
demonstrates the $\propto\eta_\star^{1/3}$ scaling [refer to
Eq.~(\ref{PK_RECONNECTION_RATE})]. The crosses and the boxes do
not follow the $\propto\eta_\star^{1/3}$ scaling for large values of
$\eta_\star$ simply because in this case the reconnection rate becomes
relatively fast and the magnetic field $B_m$ outside of the
reconnection layer is not piled up as much as in the case when
$\eta_\star$ is small and the reconnection rate is relatively slow. As
a result, our rate curves flatten at large values of $\eta_\star$. 
The central and the bottom plots on the left in
Fig.~\ref{FIG_RATES_1_3} show the observed values of coefficients
$\epsilon$ and $\gamma$, which are directly calculated from the
simulation data by using equations~(\ref{B_ENERGY})
and~(\ref{J_YY}). As we can see, $\epsilon$ and $\gamma$ are of order
of unity, as one expects. The three plots on the right in 
Fig.~\ref{FIG_RATES_1_3} are the same as the three plots on the left
except point M is now chosen as the point on the x-axis at which
$(\eta j_z)_m=(1/10)\eta_oj_o$. Comparing the plots on the left and
the plots on the right, we see that the choice of point M is indeed
important. Note that for the choice $(\eta j_z)_m=(1/3)\eta_oj_o$ for
the position of point M the half-thickness of the reconnection layer
$\delta_o$, defined as the abscissa of point M, increases from $0.011$
to $0.050$ as $\eta_\star$ increases from zero to its maximal value
shown on the plots in Fig.~\ref{FIG_RATES_1_3}. At the same time, for
the choice  $(\eta j_z)_m=(1/10)\eta_oj_o$ for the position of point M
$\delta_o$ ranges from $0.018$ to $0.056$, which is noticeably
larger. Perhaps simulations of forced reconnection with a strict
control of position of point M together with control of the outside
field $B_m$ and its scale $L$, could be better suited for comparison to
our theoretical model. Such simulations are beyond the scope of this
paper. However, see more discussion of forced reconnection in the next
section. Finally, Figure~\ref{FIG_RATES_2_3} shows the same results as
Fig.~\ref{FIG_RATES_1_3}, except the former has plots for a
smaller value of the critical current, $j_c=4.135$. 

We believe that the results presented in Figs.~\ref{FIG_RATES_1_3}
and~\ref{FIG_RATES_2_3} generally confirm our theoretical model. In
particular, in all cases the theoretical reconnection rates and the
rates observed in the simulations do not differ by more than 33\%.
The observed relatively small discrepancy between the theoretical and
simulated rates is mainly due to the coefficient $\epsilon$ not being
precisely constant in the simulations, while the variations of
coefficient $\gamma$ are somewhat less important 
[see Eq.~(\ref{RATE_GENERAL}) for the theoretical reconnection
rate]. This discrepancy can be due to two causes. First, our
theoretical model is general, but approximate, and second, the plasma
is compressible in the simulations, while it is assumed to be
incompressible in the theoretical model.


\section{Discussion and conclusions}
\label{CONCLUSIONS}

Let us summarize our main results. In this paper we take a new
theoretical approach to the calculation of the rate of
quasi-stationary magnetic reconnection. Our approach is based on
analytical derivations of the reconnection rate from the resistive MHD
equations in a small region of space that is localized about the
center of a thin reconnection layer and has its size equal to the
layer thickness. Our local-equations approach turns out to be feasible
and insightful. It allows us to consider magnetic reconnection with an
arbitrary anomalous resistivity and to calculate the reconnection rate
for this general case [see Eq.~(\ref{RATE})]. 
We find the interesting and important result that if plasma is
incompressible and reconnection is quasi-stationary, then the
reconnection rate is determined by the anomalous resistivity function
$\eta(j,x,y)$ and by the strength and structure of the magnetic field
just outside of the reconnection layer (i.e.~at point M in
Fig.~\ref{FIG_RECONNECTION_LAYER}). Thus, we find that the global
magnetic field and its configuration are not directly relevant for the
purpose of calculation of a quasi-stationary reconnection rate,
although, of course, the local magnetic field outside the reconnection
layer depends on the global field.

One of the major results of this paper is that in the case of constant
resistivity, $\eta\equiv{\rm const}=\eta_o$, the magnetic reconnection
rate is the slow Sweet-Parker reconnection rate and not the fast
Petschek reconnection rate [refer to Eqs.~(\ref{SP_RECONNECTION})]. 
This result agrees with numerical simulations and at the same time
contradicts the result of the original Petschek theoretical model. In
the framework of our theoretical approach, based on local equations,
the reason for this contradiction can be understood as follows: In the
Petschek model our parameter 
$\upsilon\equiv(\partial V_y/\partial y)_o= 
-(\partial V_x/\partial x)_o$, which is equal to the first order
partial derivatives of the incompressible plasma velocities at the
reconnection layer center, is basically treated as a free parameter. 
This is because in the Petschek model $\upsilon$ can be estimated as
the ratio of the plasma outflow velocity (equal to the Alfven velocity
for viscosity-free plasma) and the reconnection layer length, 
$\upsilon=(\partial V_y/\partial y)_o\approx V_A/L'$, and the
layer length $L'$ is treated as a free parameter by Petschek. In his 
model $L'$ is taken to be equal to the minimal possible value, that
the Petschek shocks do not seriously perturb the magnetic field in the
upstream region. This is $L'\approx(L/S_o)(\ln S_o)^2$, where
$S_o=V_AL/\eta_o$ is the Lundquist
number~\cite{petschek_1964,kulsrud_2001}. In this case 
$\upsilon\approx(V_A^2/\eta_o)(\ln S_o)^{-2}$ and, according to
equations~(\ref{V_R}) and~(\ref{UPSILON}), the reconnection velocity
in this case is equal to $V_R\approx V_A(\ln S_o)^{-1}$, which is the
Petschek result. On the other hand, in our theoretical model the
parameter $\upsilon$ is not treated as a free parameter. In fact, our
three physical parameters $\upsilon\equiv(\partial V_y/\partial y)_o$,
$\beta\equiv(\partial B_x/\partial y)_o$ and $j_o\equiv j_z(x=0,y=0)$
are connected to each other and must be calculated from 
equations~(\ref{B_ENERGY}),~(\ref{SECOND_MHD}) and~(\ref{THIRD_MHD}). 
Let us discuss the meaning of these
equations. Equation~(\ref{B_ENERGY}) is the equation of magnetic
energy conservation. It says that the rate of supply of the magnetic
field energy into the reconnection layer $\upsilon B_m^2$ must be
equal to the rate of the resistive dissipation of this energy 
$\eta_oj_o^2$. In the Petschek model $\upsilon$ and, accordingly, the
rate of the magnetic energy supply $\upsilon B_m^2$ are basically
prescribed by hand (resulting in an {\it ad hoc} fast reconnection),
while in our model they are self-consistently calculated from the MHD
equations. Equation~(\ref{SECOND_MHD}) is the equation of plasma
acceleration along the reconnection layer. It says that the magnetic
tension force $\beta j_o$ must be large enough in order to be able to
push out all the plasma along the layer that is supplied into the
layer. Finally, equation~(\ref{THIRD_MHD}) is the equation of 
spatial homogeneity of the electric field z-component along the
reconnection layer. This equation sets an upper limit on the product
$\upsilon\beta$ in the case of a quasi-stationary reconnection and is
directly related to the calculations and arguments given by
Kulsrud~\cite{kulsrud_2001} in the framework of the global-equations
theoretical approach, see Ref.~\footnote{ 
The spatial homogeneity of the electric field z-component and the
resulting upper limit on $\upsilon\beta$ are directly related to the
explanation of why the Petschek reconnection model does not work, as
shown by Kulsrud~\cite{kulsrud_2001} in the framework of the
global-equations. Kulsrud's argument is that the length of Petschek
reconnection layer $L'$ is not a free parameter, but must be
determined by the condition that the perpendicular magnetic field
component $B_x$ has to be regenerated by  the rotation of the parallel
field component $B_y$ at the same rate as it is being swept away by
the downstream flow. It is easy to see that the integration of
``local'' equation~(\ref{MHD_EQUATION}) without the resistivity term
over the area of the contour  
O$\to$M$\to$${\tilde{\rm M}}$$\to$${\tilde{\rm O}}$ shown in
Fig.~\ref{FIG_RECONNECTION_LAYER} will result in the same ``global''
equation for the balance of the $B_x$ field, which was used by Kulsrud
in his work~\cite{kulsrud_2001}.
}.
As a result, none of parameters $\upsilon$, $\beta$ and $j_o$ can be
treated as free parameters, and all of them must be self-consistently
determined from the MHD equations. 

It is very instructive to briefly examine our results from the two
distinct points of view in connection with the reconnection problem,
which are expressed in numerous papers on computer simulation of
magnetic reconnection. These two points of view are: unforced (free)
magnetic reconnection and forced magnetic reconnection. In the
case of unforced reconnection, one should solve our main
equation~(\ref{RATE}) for the current $j_o$ at the reconnection layer
center and then solve for the reconnection velocity $V_R$ by using
equation~(\ref{V_R}). The solution for $j_o$ and $V_R$ will depend on
$B_m$, which is the strength of the magnetic field outside the
reconnection layer and which enters the right-hand-side of
equation~(\ref{RATE}). On the other hand, in the case of forced
magnetic reconnection the reconnection velocity $V_R$ is prescribed
and fixed. In this case the magnetic field outside the reconnection 
layer $B_m$ should be treated as an unknown quantity, and
equations~(\ref{RATE}) and~(\ref{V_R}) should be solved together in
order to find the correct quasi-stationary values of $j_o$ and
$B_m$. In other words, in the forced reconnection case an initially
weak outside magnetic field $B_m$ gets piled up to higher and higher
values until the resulting current $j_o$ in the reconnection layer
becomes large enough to be able to exactly match the prescribed
reconnection velocity $V_R$ and to be able to reconnect all magnetic
flux and magnetic energy, which are supplied into the reconnection
region in the quasi-stationary reconnection regime. 

Finally, a couple of words about plasma viscosity and guide field
and their effect on magnetic reconnection. First, according to our
equations~(\ref{RATE}) and~(\ref{V_R}), in the case when the
resistivity is constant, $\eta\equiv{\rm const}=\eta_o$, and the
plasma viscosity is much larger than resistivity, $\nu\gg\eta_o$,
the reconnection velocity becomes 
$V_R/V_A\approx S_o^{-1/2}(\eta_o/\nu)^{1/4}$, which is 
$\sqrt[4]{\nu/\eta_o}$ times smaller than the Sweet-Parker
reconnection velocity given by
formula~(\ref{SP_RECONNECTION}), see~\cite{park_1984}. Thus, we see
that the reconnection rate becomes smaller when the plasma viscosity
becomes large. However, note that in many astrophysical and laboratory
applications plasmas are very hot and highly rarefied. Under these
conditions the ion gyro-radius becomes much shorter than the ion
mean-free-path, and the plasma becomes strongly magnetized. As a
result, the plasma viscosity becomes the Braginskii viscosity, which
is dominated by magnetized ions~\cite{braginskii_1965}. In this case
in all our equations above the isotropic viscosity $\nu$, which is
proportional to the ion mean-free-path, should be replaced by the
Braginskii perpendicular viscosity, which is proportional to the ion
gyro-radius and is much smaller than the perpendicular viscosity in a
strongly magnetized plasma. 
Second, according to our results, in two-and-a-half dimensional
geometry the guide field $B_z$ has no effect on the quasi-stationary
reconnection rate. Indeed, in our derivations the guide field appears
only as magnetic pressure $B_z^2/2$ term in addition to the plasma
pressure. The combined pressure $P$ enters
equations~(\ref{V_Y_EQUATION}) and~(\ref{V_Y_EQUATION_2}) of plasma
acceleration along the reconnection layer and the value of spatial
derivative of $P$ is given by equation~(\ref{PRESSURE_TERM}), which
does not involve $B_z$. Thus, the guide field gets eliminated and does
not enter into our final equation~(\ref{RATE}) for the reconnection
rate. However, if one assumes that the anomalous resistivity $\eta$
depends on x- and y-components of the current 
$j_x=\partial B_z/\partial y$ and $j_y=-\partial B_z/\partial x$ in
addition to its dependence on the z-component of the current $j_z$,
then the reconnection rate will depend on the guide field $B_z$. 

In this paper we consider quasi-stationary magnetic reconnection in a
thin reconnection layer. We leave a study of tearing modes instability
in a reconnection layer and non-quasi-stationary reconnection for
a future paper.


\begin{acknowledgments}
It is our special pleasure to thank Ellen Zweibel and Dmitri Uzdensky
for many stimulating discussions and useful comments. We are 
grateful to Zoran Mikic and Samuel Vainshtein for suggesting to us 
the convenient computational setup used in our simulations. We would
also like to thank Andrey Beresnyak, Amitava Bhattacharjee, Stas
Boldyrev, Fausto Cattaneo, Jeremy Goodman, Hantao Ji, Alexander
Obabko, Robert Rosner and Masaaki Yamada for a number of valuable
comments.
This work was supported by the Center for Magnetic Self-Organization
(CMSO) grant. The numerical simulations were supported by a
Center for Magnetic Reconnection Studies (CMRS) grant. The software
used in this work was in part developed by the DOE supported
ASC/Alliances Center for Astrophysical Thermonuclear Flashes at the
University of Chicago. The simulations were carried out on DOE
computers at the Oak Ridge National Laboratory.
\end{acknowledgments}


\appendix

\section{Derivation of equations~(\ref{PRESSURE_TERM})
and~(\ref{VISCOSITY_TERM})} 
\label{APPENDIX_A}

Below, for brevity, we assume that spatial derivatives are
to be taken with respect to all indexes that are listed after the
comma signs in the subscripts, 
e.g.~$V_{x,yy}\equiv\partial^2 V_x/\partial y^2$.

We derive equation~(\ref{PRESSURE_TERM}) first. The derivation is
somewhat analogous to the Sweet-Parker derivation of the pressure
decrease along the reconnection layer, which leads to
equation~(\ref{SP_ENERGY}). Namely, to find the pressure decrease and
the pressure second derivative along the layer, we integrate the
pressure gradient vector along the contour 
O$\to$M$\to$${\tilde{\rm M}}$$\to$${\tilde{\rm O}}$ shown in
Fig.~\ref{FIG_RECONNECTION_LAYER} and use the force balance
condition for the plasma across the reconnection layer. Now we
carry out these calculations in a mathematically precise way. 

For infinitesimally small values of the y-coordinate, taking into
account the symmetry of the reconnection layer with respect to the x-
and y-axes and plasma incompressibility, we use the following
Taylor expansions in $y$ for the plasma velocity ${\bf V}(x,y)$ and
for the magnetic field ${\bf B}(x,y)$: 
\beq
\begin{array}{l}
V_x=V_x^{(0)}(x)+(y^2/2)V_{x,yy}^{(0)}(x)+(y^4/24)V_{x,yyyy}^{(0)}(x),
\\
B_x=yB_{x,y}^{(0)}(x)+(y^3/6)B_{x,yyy}^{(0)}(x),
\\
V_y = -yV_{x,x}^{(0)}(x)+(y^3/6)V_{y,yyy}^{(0)}(x),
\\
B_y = B_y^{(0)}(x)+(y^2/2)B_{y,yy}^{(0)}(x),
\end{array}
\label{APP_A_EXPANSIONS}
\eeq
where the variables with the superscripts $\,^{(0)}$ are taken at $y=0$ and
depend only on $x$. Assuming quasi-stationarity of reconnection
($\partial/\partial t=0$) and plasma incompressibility, the MHD
equation for the plasma velocity ${\bf V}$ is 
\beq
{\bf\nabla}P+{\bf\nabla}(B_x^2+B_y^2)/2&=&
-\rho({\bf V}{\bf\nabla}){\bf V}+
({\bf B}{\bf\nabla}){\bf B}
\nonumber\\
&&+\rho\nu{\bf\nabla}^2{\bf V},
\label{APP_A_V_EQUATION}
\eeq
where $P$ is the sum of the plasma pressure and the z-component
field magnetic pressure $B_z^2/2$. Let us calculate line integrals of
the left- and right-hand-sides of equation~(\ref{APP_A_V_EQUATION})
along the contour O$\to$M$\to$${\tilde{\rm M}}$$\to$${\tilde{\rm O}}$
shown in Fig.~\ref{FIG_RECONNECTION_LAYER}. First, the line integral
of the left-hand-side is obviously
\beq
&&P(0,{\tilde y})-P(0,0)+(1/2)B_x^2(0,{\tilde y})
\nonumber\\
&&\qquad\qquad\qquad = ({\tilde y}^2/2)[P_{,yy}(0,0)+\beta^2],
\qquad
\label{APP_A_LHS_INTEGRAL}
\eeq
where ${\tilde y}$ is the y-coordinate of points ${\tilde{\rm M}}$ and
${\tilde{\rm O}}$, and we use the formulas $B_y\equiv0$ on the y-axis,
$B_x=0$ at point O and $B_x(0,y)=\beta y$ for small $y$ [see the
definition of $\beta$ in Eq.~(\ref{SECOND_MHD})]. 
Second, using expansion formulas~(\ref{APP_A_EXPANSIONS}), we
calculate the following line integrals along the contour 
O$\to$M$\to$${\tilde{\rm M}}$$\to$${\tilde{\rm O}}$,
up to the second order in ${\tilde y}$:
\beq
\int[\rho({\bf V}{\bf\nabla}){\bf V}]{\bf dl} &=&
\rho\frac{{\tilde y}^2}{2}
\Bigl(\bigl[V_{x,x}^{(m)}\bigr]^2-
V_x^{(m)}\bigl[V_{x,xx}^{(m)}+V_{x,yy}^{(m)}\bigr]
\nonumber\\
&&{}+2\int_O^M V_{x,x}^{(0)}V_{x,yy}^{(0)}\,dx\Bigr),
\label{APP_A_V_INTEGRAL}
\\
\int[({\bf B}{\bf\nabla}){\bf B}]{\bf dl} &=&
\frac{{\tilde y}^2}{2}
\Bigl(B_{y}^{(m)}B_{y,yy}^{(m)}+\beta^2+
B_{x,y}^{(m)}j_{z}^{(m)}
\nonumber\\
&&-\int_O^M \bigl[B_{y}^{(0)}B_{x,yyy}^{(0)}
\nonumber\\
&&\qquad{}+B_{x,y}^{(0)}B_{y,yy}^{(0)}\bigr]dx\,\Bigr),
\label{APP_A_B_INTEGRAL}
\\
\int[\rho\nu{\bf\nabla}^2{\bf V}]{\bf dl} &=&
\rho\nu\frac{{\tilde y}^2}{2}
\Bigl(2V_{y,yyy}^{(m)}-V_{x,xxx}^{(m)}-V_{y,yyy}^{(0)}(0)
\nonumber\\
&&-\int_O^M V_{x,yyyy}^{(0)}\,dx\Bigr),
\label{APP_A_NU_INTEGRAL}
\eeq
where the variables with the superscripts $\,^{(m)}$ are calculated at
point M (see Fig.~\ref{FIG_RECONNECTION_LAYER}). 

Next we estimate the terms on the right-hand-sides of
Eqs.~(\ref{APP_A_V_INTEGRAL})-(\ref{APP_A_NU_INTEGRAL}).
Recall the notations $L'$ and $\delta_o$ for the half-length
and half-thickness of the reconnection layer (see
Fig.~\ref{FIG_RECONNECTION_LAYER}), and that 
$\delta_o/L'\ll1$ (which is our assumption of a thin reconnection
layer). The z-current inside the reconnection layer is approximately
equal to $j_o\approx B_m/\delta_o$ [see Eq.~(\ref{J_O})], where
$B_m=B_{y}^{(m)}$ is the field at point M, while the z-current outside
the layer is $j_{z}^{(m)}\lesssim B_m/L'$. The last two terms on the
right-hand-side of equation~(\ref{APP_A_B_INTEGRAL}) can be estimated
as $B_{x,y}^{(m)}j_{z}^{(m)}\sim \beta B_m/L'\sim
\beta j_o(\delta_o/L')\ll\beta j_o$ and 
$\int_O^M [B_{y}^{(0)}B_{x,yyy}^{(0)}+B_{x,y}^{(0)}B_{y,yy}^{(0)}]dx
\sim\delta_o B_m\beta/L'^2\sim\beta j_o(\delta_o/L')^2\ll
\beta j_o$. Next, from equation~(\ref{MHD_EQUATION}) with the
resistivity term dropped we see that outside the reconnection
layer the typical scale of the plasma velocity ${\bf V}$ is about the
same as the typical scale of the magnetic field ${\bf B}$ and,
therefore, can not be smaller than $L'$. Thus, the estimates of the
three terms on the right-hand-side of
equation~(\ref{APP_A_V_INTEGRAL}) are 
$[V_{x,x}^{(m)}]^2\sim V_x^{(m)}[V_{x,xx}^{(m)}+V_{x,yy}^{(m)}]\sim
V_R^2/L'^2\sim\upsilon^2(\delta_o/L')^2\ll \upsilon^2$ [see
Eq.~(\ref{UPSILON_ESTIMATE})] and
$\int_O^M V_{x,x}^{(0)}V_{x,yy}^{(0)}\,dx\sim\delta_o\upsilon V_R/L'^2
\sim \upsilon^2(\delta_o/L')^2\ll \upsilon^2$. The estimates of the
four terms on the right-hand-side of
equation~(\ref{APP_A_NU_INTEGRAL}) are 
$V_{y,yyy}^{(m)}\sim V_{x,xxx}^{(m)}\sim V_R/L'^3\sim
\upsilon\delta_o/L'^3\ll\upsilon/\delta_o^2$, 
$V_{y,yyy}^{(0)}(0)\sim\upsilon/L'^2\ll\upsilon/\delta_o^2$ and 
$\int_O^M V_{x,yyyy}^{(0)}\,dx\sim\delta_oV_R/L'^4\sim
\upsilon\delta_o^2/L'^4\ll\upsilon/\delta_o^2$.
Note that here we use $L'$ for estimation of
y-derivatives. In fact, using the global scale $L$ of the magnetic
field outside the reconnection layer would have been more appropriate
for some of the estimations (as shown in
Appendix~\ref{APPENDIX_B}). However, using $1/L'\ge1/L$ for
the upper estimates of the $\partial/\partial y$ derivatives is
perfectly fine for the purposes in this appendix. 

Next, calculating the line integral of the right-hand-side of
Eq.~(\ref{APP_A_V_EQUATION}) by using
formulas~(\ref{APP_A_V_INTEGRAL})-(\ref{APP_A_NU_INTEGRAL}) and
our estimates made in the previous paragraph, and then equating the
result to the right-hand-side of Eq.~(\ref{APP_A_LHS_INTEGRAL}), we
easily obtain equation~(\ref{PRESSURE_TERM}). Note that the first term
on the right-hand side of equation~(\ref{PRESSURE_TERM}) can also be
written in terms of the second derivative of the magnetic pressure at
point M: $B_m(\partial^2 B_y/\partial y^2)_m=
(\partial^2/\partial y^2)(B_y^2/2+B_x^2/2)_m-\beta^2=
(\partial^2/\partial y^2)(B_y^2/2+B_x^2/2)_m+{\rm o}\{\beta j_o\}$
(note that $\beta\ll j_o$ because the reconnection layer is
thin). Therefore, as noted by Zweibel~\cite{zweibel_2005},
equation~(\ref{SECOND_MHD}) is similar to Bernoulli's equation.  

Now we derive equation~(\ref{VISCOSITY_TERM}), which gives an 
approximate estimate of the viscosity term
$\rho\nu\left[{\bf\nabla}^2(\partial V_y/\partial y)\right]_o$ 
in equation~(\ref{V_Y_EQUATION_2}). 
We make this estimate as follows: Note that $V_{y,yx}=0$ on the
y-axis because of the symmetry of the problem relative to this
axis. 
Therefore, from the second order Taylor expansion of $V_{y,y}(x,0)$
in $x$, we obtain an approximate formula 
$V_{y,yxx}(0,0)\approx[V_{y,y}(\delta_o,0)-V_{y,y}(0,0)]/\delta_o^2
\approx-V_{y,y}(0,0)/\delta_o^2=-\upsilon/\delta_o^2$, where 
we take into account that
$V_{y,y}(\delta_o,0)\sim V_R/L'\approx\upsilon\delta_o/L'\ll\upsilon$.
We can rewrite this approximate formula for $V_{y,yxx}(0,0)$ as the
following exact formula: 
$V_{y,yxx}(0,0)=-C\upsilon/\delta_o^2$, where $C$ is an unknown
coefficient of order unity. Our numerical simulations of reconnection
with constant resistivity show that $C$ is indeed about unity if
$\delta_o$ is estimated by equation~(\ref{J_O}). Thus we
immediately find that 
$\rho\nu\left[{\bf\nabla}^2(\partial V_y/\partial y)\right]_o=
\rho\nu[V_{y,yxx}(0,0)+V_{y,yyy}(0,0)]\approx
\rho\nu V_{y,yxx}(0,0)\approx-\rho\nu\upsilon/\delta_o^2$,
which is equation~(\ref{VISCOSITY_TERM}). Here we also use
$V_{y,yyy}(0,0)\sim\upsilon/L'^2\ll\upsilon/\delta_o^2$.


\section{Derivation of equation~(\ref{J_YY})}
\label{APPENDIX_B}

Here as in Appendix~\ref{APPENDIX_A}, we assume that spatial
derivatives are to be taken with respect to all indexes that are
listed after the comma signs in the subscripts, 
e.g.~$B_{x,xy}\equiv\partial^2B_x/\partial x\partial y$. We derive
equation~(\ref{J_YY}) in two steps.

First, we estimate $B_{x,yyy}$ at points O and M (see
Fig.~\ref{FIG_RECONNECTION_LAYER}), since we will need these estimates
below. Consider the formula $B_{x,x}=-B_{y,y}$, which represents the
fact that the magnetic field is divergence-free. Take the
$\partial/\partial y$ and $\partial^3/\partial y^3$ derivatives of
this formula and integrate the resulting equations along the interval
OM shown in Fig.~\ref{FIG_RECONNECTION_LAYER}. We obtain
\beq
B_{x,y}^{(m)}-\beta &=& \int_O^M\!\! B_{x,yx}\,dx
=-\int_O^M\!\! B_{y,yy}\,dx
\nonumber\\
{}&\approx&-\delta_o B_{y,yy}^{(m)},
\label{APP_B_Bxy}
\\
B_{x,yyy}^{(m)}-B_{x,yyy}^{(o)} &=& \int_O^M\!\! B_{x,yyyx}\,dx
=-\int_O^M\!\! B_{y,yyyy}\,dx
\nonumber\\
{}&\approx&-\delta_o B_{y,yyyy}^{(m)},
\label{APP_B_Bxyyy}
\eeq
where the variables with the superscripts $\,^{(o)}$ and $\,^{(m)}$
are taken at points O and M respectively and $\beta=B_{x,y}^{(o)}$
[see Eq.~(\ref{SECOND_MHD})]. Note that $\delta_o$ is the
half-thickness of the reconnection layer, equal to the abscissa of
point M (see Fig.~\ref{FIG_RECONNECTION_LAYER}). In making the
estimates of the integrals on the right-hand-sides of
Eqs.~(\ref{APP_B_Bxy}) and~(\ref{APP_B_Bxyyy}), we take into
account that $B_{y,yy}^{(o)}=B_{y,yyyy}^{(o)}=0$ because $B_y\equiv0$
on the y-axis. Now, using Eq.~(\ref{APP_B_Bxy}), we estimate that
$B_{x,y}^{(m)}\approx\beta-\delta_o B_{y,yy}^{(m)}$. Let $L$ be the
global scale of the magnetic field outside the reconnection
layer. Then, since point M is located outside the reconnection layer,
we have $B_{y,yyyy}^{(m)}\sim \pm B_{y,yy}^{(m)}/L^2$ and 
$B_{x,yyy}^{(m)}\sim \pm B_{x,y}^{(m)}/L^2$. Next, using these
estimates, the above estimate for $B_{x,y}^{(m)}$ and
Eq.~(\ref{APP_B_Bxyyy}), we obtain the following formula:
\beq
B_{x,yyy}^{(o)}\sim B_{x,yyy}^{(m)}\sim
\pm\beta/L^2\pm(\delta_o/L^2)B_{y,yy}^{(m)}.
\label{APP_B_Bx_ESTIMATES}
\eeq

Second, we estimate $j_{z,yy}$ at point O. Consider Ampere's law
formula $j_z=B_{y,x}-B_{x,y}$. We take the $\partial^2/\partial y^2$
derivative of this equation and integrate the result along the
interval OM shown in Fig.~\ref{FIG_RECONNECTION_LAYER}. We find that
\beq
\int_O^Mj_{z,yy}\,dx &=&
\int_O^MB_{y,yyx}\,dx-\int_O^MB_{x,yyy}\,dx
\nonumber\\
{}&=& B_{y,yy}^{(m)}-\int_O^MB_{x,yyy}\,dx.
\eeq
The integral $\int_O^MB_{x,yyy}\,dx$ can be estimated by using
formula~(\ref{APP_B_Bx_ESTIMATES}). The integral of $j_{z,yy}$ can 
be estimated as $\int_O^Mj_{z,yy}\,dx\sim\delta_o j_{z,yy}^{(o)}$.
As a result, we obtain
\beq
\delta_o j_{z,yy}^{(o)}\approx 
B_{y,yy}^{(m)}\pm\delta_o\frac{\beta}{L^2}
\pm\delta_o\frac{\delta_o}{L^2}B_{y,yy}^{(m)}\approx B_{y,yy}^{(m)},
\label{APP_B_APP_B_Jzyy}
\eeq
where we use $\delta_o\ll L$ and 
$\delta_o\beta/L^2=(\delta_o j_o/L^2)(\beta/j_o)\approx
(B_m/L^2)(\beta/j_o)\approx \pm B_{y,yy}^{(m)}(\beta/j_o)\ll
|B_{y,yy}^{(m)}|$, see Eq.~(\ref{J_O}) and note that $\beta\ll j_o$
because reconnection layer is thin. 
Finally, substituting an estimate $\delta_o\approx B_m/j_o$ into
formula~(\ref{APP_B_APP_B_Jzyy}), we obtain equation~(\ref{J_YY}) 
with coefficient $\gamma\approx 1$.

As suggested by Uzdensky~\cite{uzdensky_2005}, there
exists a nice graphical interpretation of the fact that the y-scale of
the current $j_z$ is about the same as the scale of the outside
magnetic field, i.e.~$j_o/(\partial^2 j_z/\partial y^2)_o\approx 
B_m/(\partial^2 B_y/\partial y^2)_m$ and that equation~(\ref{J_YY})
holds. There can be two different cases of the reconnection layer
geometry. First, the half-length of the reconnection layer $L'$ can be
approximately equal to the global scale $L$ of the outside field. In
this case the reconnection is Sweet-Parker and $L'\approx L$ is the
only available scale in the y-direction. Therefore, in this case 
$j_o/|\partial^2 j_z/\partial y^2|_o\approx 
B_m/|\partial^2 B_y/\partial y^2|_m\approx L^2\approx L'^2$.
In the second case the reconnection layer half-length is much smaller
than the global scale, $L'\ll L$, and the reconnection is fast 
(relative to the Sweet-Parker reconnection). In this case,
the z-current $j_z(0,y)$ on the y-axis drops abruptly, as the
y-coordinate passes value $L'$ and point $(0,y)$ moves from the region
inside the reconnection layer to the region of the outflowing plasma
that is located between the Petschek shocks (see
Fig.~\ref{FIG_RECONNECTION_LAYER}). However, the z-current $j_z$
stays large inside the shocks. In other words, $j_z$ is a smooth
function (on the global scale $L$) along the lines that lie inside the
reconnection layer and extend into the shock separatrices. Thus, in
this case, despite $L'\ll L$, the y-scale of the z-current at the
reconnection layer central point O is still $L$. This graphical
interpretation is well demonstrated by the bottom-left plot of the
current for the Petschek-Kulsrud (``P-K'') reconnection case in
Fig.~\ref{FIG_ROSETTE}.



\begin{thebibliography}{99}

\bibitem{sweet_1958}
P. A. Sweet, in {\it Electromagnetic Phenomena in Ionized Gases},
edited by B. Lehnert (Cambridge University Press, New York, 1958),
p. 123.

\bibitem{parker_1963}
E. N. Parker, Astrophys. J., Suppl. Ser. {\bf 8}, 177 (1963).

\bibitem{petschek_1964}
H. E. Petschek, in {\it AAS-NASA Symposium on Solar Flares} NASA SP50
(National Aeronautics and Space Administration, 
Washington, DC, 1964), p. 425.

\bibitem{biskamp_1986}
D. Biskamp, Phys. Fluids {\bf 29}, 1520 (1986).

\bibitem{kulsrud_2001}
R. M. Kulsrud, Earth, Planets and Space {\bf 53}, 417 (2001);
astro-ph/0007075.

\bibitem{uzdensky_2000}
D. A. Uzdensky and R. M. Kulsrud, Phys. Plasmas {\bf 7}, 4018 (2000).

\bibitem{ugai_1977}
M. Ugai and T. Tsuda, J. Plasma Phys. {\bf 17}, 337 (1977).

\bibitem{tsuda_1977}
T. Tsuda and M. Ugai, J. Plasma Phys. {\bf 18}, 451 (1977).

\bibitem{hayashi_1978}
T. Hayashi and T. Sato, J. Geophys. Res. {\bf 83}, 217 (1978).

\bibitem{sato_1979}
T. Sato and T. Hayashi, Phys. Fluids {\bf 22}, 1189 (1979).

\bibitem{scholer_1989}
M. Scholer, J. Geophys. Res. {\bf 94}, 8805 (1989).

\bibitem{lazarian_1999}
A. Lazarian and E. T. Vishniac, Astrophys. J. {\bf 517}, 700 (1999).

\bibitem{kim_2001}
E. Kim and P. H. Diamond, Astrophys. J. {\bf 556}, 1052 (2001)

\bibitem{biskamp_1995}
D. Biskamp, E. Schwarz, and J. F. Drake, 
Phys. Rev. Lett. {\bf 75}, 3850 (1995).

\bibitem{bhattacharjee_2003}
A. Bhattacharjee, Z. W. Ma, and X. Wang, 2003, in 
{\it Turbulence and Magnetic Fields in Astrophysics}, 
edited by E. Falgarone and T. Passot (Springer, 2003), 
Lecture Notes in Physics {\bf 614}, p. 351.

\bibitem{craig_2005}
I. J. D. Craig and P. G. Watson, Phys. Plasmas {\bf 12}, 012306
(2005).

\bibitem{drake_2003}
J. F. Drake, M. Swisdak, C. Cattell, M. A. Shay, B. N. Rogers, 
and A. Zeiler, Science {\bf 299}, 873 (2003).

\bibitem{hanasz_2003}
M. Hanasz and H. Lesch, Astron. Astrophys. {\bf 404}, 389 (2003).

\bibitem{heitsch_2003}
F. Heitsch and E. G. Zweibel, Astrophys. J. {\bf 583}, 229 (2003).

\bibitem{rogers_2003}
B. N. Rogers, R. E. Denton, and J. F. Drake, 
J. Geophys. Res. {\bf 108}, 1111 (2003).

\bibitem{shay_2001}
M. Shay, J. Drake, B. Rogers, and R. Denton, 
J. Geophys. Res. {\bf 106}, 3759 (2001).

\bibitem{kulsrud_2005}
R. M. Kulsrud, {\it Plasma Physics for Astrophysics} 
(Princeton University Press, 2005).

\bibitem{furth_1963}
H. P. Furth, J. Killeen, and M. N. Rosenbluth, 
Phys. Fluids {\bf 6}, 459 (1963).

\bibitem{landau_1983} 
L. D. Landau and E. M. Lifshitz,
{\it Electrodynamics of continuous media} (New York, Pergamon, 1983).

\bibitem{vasyliunas_1975}
V. M. Vasyliunas, Rev. Geophys. Space Phys. {\bf 18}, 303 (1975). 

\bibitem{park_1984}
W. Park, D. A. Monticello, and R. B. White, 
Phys. Fluids {\bf 27}, 137 (1984).

\bibitem{breslau_2003}
J. A. Breslau and S. C. Jardin, 
Physics of Plasmas {\bf 10}, 1291 (2003).

\bibitem{biskamp_2001}
D. Biskamp and E. Schwarz, Phys. Plasmas {\bf 8}, 4729 (2001).

\bibitem{mikic_2003}
Z. Mikic and S. Vainshtein, unpublished (2003).

\bibitem{calder_2002}
A. C. Calder, \etal, 
Astrophys. J., Suppl. Ser. {\bf 143}, 201 (2002).

\bibitem{fryxell_2000}
B. Fryxell, \etal, Astrophys. J., Suppl. Ser. {\bf 131}, 273 (2000).

\bibitem{braginskii_1965}
S. I. Braginskii, Rev. Plas. Phys. {\bf 1}, 205 (1965).

\bibitem{zweibel_2005}
E. G. Zweibel, private communication (2005).

\bibitem{uzdensky_2005}
D. A. Uzdensky, private communication (2005).

\bibitem{harris_1962}
E. G. Harris, Nuovo Cimento {\bf 23}, 115 (1962).

\end{thebibliography}
\end{document}